\documentclass[twocolumn,preprintnumbers,amsmath,amssymb,aps,prb,longbibliography]{revtex4-1}

\usepackage{graphicx}
\usepackage{calrsfs}

\begin{document}

\title{Shapiro Steps and Stability of Skyrmions Interacting with Alternating Anisotropy Under the Influence of ac and dc Drives}
\author{J. C. Bellizotti Souza$^{1}$,
            N. P. Vizarim$^{2}$, 
            C. J. O. Reichhardt$^3$, 
            C. Reichhardt$^3$
            and P. A. Venegas$^4$}
            
\affiliation{$^1$ POSMAT - Programa de P\'os-Gradua\c{c}\~ao em Ci\^encia e Tecnologia de Materiais, S\~ao Paulo State University (UNESP), School of Sciences, Bauru 17033-360, SP, Brazil}

\affiliation{$^2$ Departament of Electronics and Telecommunications Engineering, S\~ao Paulo State University (UNESP), School of Engineering, S\~ao J\~oao da Boa Vista 13876-750, SP, Brazil}

\affiliation{$^3$ Theoretical Division and Center for Nonlinear Studies, Los Alamos National Laboratory, Los Alamos, New Mexico 87545, USA}

\affiliation{$^4$ Department of Physics, S\~ao Paulo State University (UNESP), School of Sciences, Bauru 17033-360, SP, Brazil}

\date{\today}

\begin{abstract}
 We use atomistic simulations to examine the sliding dynamics of a skyrmion in a two-dimensional system containing a periodic one-dimensional stripe pattern of variations between low and high values of the perpendicular magnetic anisotropy. The skyrmion changes in size as it crosses the interface between two anisotropy regions. Upon applying combined dc and ac driving in either parallel or perpendicular directions, we observe a wide variety of Shapiro steps, Shapiro spikes, and phase-locking phenomena. The phase-locked orbits have two-dimensional dynamics due to the gyrotropic or Magnus dynamics of the skyrmions, and are distinct from the phase-locked orbits found for strictly overdamped systems. Along a given Shapiro step when the ac drive is perpendicular to the dc drive, the velocity parallel to the ac drive is locked while the velocity in the perpendicular direction increases with increasing drive to form Shapiro spikes. At the transition between adjacent Shapiro steps, the parallel velocity jumps up to the next step value, and the perpendicular velocity drops. The skyrmion Hall angle shows a series of spikes as a function of increasing dc drive, where the jumps correspond to the transition between different phase-locked steps. At high drives, the Shapiro steps and Shapiro spikes are lost. When both the ac and dc drives are parallel to the stripe periodicity direction, Shapiro steps appear, while if the dc drive is parallel to the stripe periodicity direction and the ac drive is perpendicular to the stripe periodicity, there are only two locked phases, and the skyrmion motion consists of a combination of sliding along the interfaces between the two anisotropy values and jumping across the interfaces.
\end{abstract}

\maketitle

\section{Introduction}

Complex systems in the presence of multiple interacting frequencies
exhibit
a range of nonlinear dynamical phenomena, 
including synchronization and phase locking \cite{pikovsky_synchronization_2003,ott_chaos_2002}.
These phenomena appear across diverse domains, from interconnected pendula \cite{bennett_huygenss_2002} 
to biological systems \cite{glass_synchronization_2001}.
A fundamental illustration of phase locking occurs in overdamped particle systems traversing a periodic 
substrate under simultaneous direct (dc) and alternating (ac) driving forces. In this case, 
resonances can occur between the ac driving 
frequency and the oscillation frequency induced by the particle motion across the substrate. 
These resonant interactions result in distinct steps in the velocity-force curves, 
since the particle remains locked to a specific velocity range over a range of external drive intervals in order to sustain the resonant state.
These resonant steps were initially observed
in Josephson junctions, 
where the current-voltage response exhibits what are called
Shapiro steps \cite{shapiro_josephson_1963,barone_physics_1982}.
Many systems demonstrating phase locking can be modeled as effectively
one-dimensional (1D) in nature, such as
Josephson junction arrays \cite{benz_fractional_1990},
incommensurate sliding charge density waves \cite{coppersmith_interference_1986,gruner_dynamics_1988,brown_subharmonic_1984},
superconducting vortices on 1D \cite{martinoli_c_1975,martinoli_static_1978,dobrovolskiy_ac_2015}
and two-dimensional (2D) periodic substrates \cite{van_look_shapiro_1999,reichhardt_phase-locking_2000}, 
driven Frenkel-Kontorova models \cite{sokolovic_devils_2017}, 
frictional systems \cite{tekic_frequency_2011}, and
colloids navigating 1D periodic substrates \cite{juniper_microscopic_2015,brazda_experimental_2017,abbott_transport_2019}.
Even within one-dimensional systems, several additional phenomena, such as fractional locking,
can arise when additional nonlinear effects are considered.

For the case of particles moving over a 2D periodic substrate, many of the same types of phase locking effects observed in 1D systems can occur;
however, the additional degrees of freedom in the 2D environment make it
possible to observe new types of effects by, for example,
applying the ac drive in a direction that is
perpendicular to the dc drive.
Such a
driving configuration produces transverse phase locking that is
distinct from Shapiro steps.
The widths of the transverse locking
steps generally increase with increasing ac amplitude
\cite{reichhardt_phase-locking_2001,marconi_transverse_2003},
unlike the oscillatory behavior observed for Shapiro steps.
For 2D substrates, it is also possible to apply biharmonic ac driving with
components that are both parallel and perpendicular to the dc drive in order
to induce circular motion of the driven particles. 
In this scenario, increasing the dc drive intensity leads to chiral scattering effects, resulting in phase-locked regions
where the particles move both parallel and perpendicular to the dc
driving direction
\cite{reichhardt_rectification_2002,reichhardt_absolute_2003}.

In most of the previously mentioned systems, the dynamics are predominantly overdamped;
however, under certain circumstances, nondissipative effects such as inertia
become important \cite{tekic_inertial_2019}.
Another possible nondissipative
term is produced by gyro-coupling or the Magnus force,
which generates velocity components perpendicular
to the forces acting on the particle.
In a 1D system, the Magnus force remains negligible, but
in 2D systems, it can significantly alter the dynamics.
An important and recent example of a system in which
the gyromagnetic forces are dominant are magnetic skyrmions
in ferromagnets. Magnetic skyrmions are topologically protected
magnetic textures \cite{nagaosa_topological_2013, je_direct_2020} 
that exhibit many similarities to overdamped particles.
Skyrmions can minimize their repulsive interactions by forming
a triangular lattice, can be set in motion by the application of
external drives, and can interact with material defects in
several different ways \cite{olson_reichhardt_comparing_2014,reichhardt_depinning_2016,reichhardt_statics_2022}.
Under external driving
in a clean sample, skyrmions move at an angle to the applied drive that is
known as the intrinsic skyrmion Hall angle
$\theta^\text{int}_\text{sk}$ \cite{nagaosa_topological_2013, litzius_skyrmion_2017, iwasaki_universal_2013, jiang_direct_2017, lin_driven_2013, lin_particle_2013}
with respect to the external drive. The sign of this angle
depends on the skyrmion winding number $Q$
\cite{nagaosa_topological_2013, litzius_skyrmion_2017, iwasaki_universal_2013, jiang_direct_2017, lin_driven_2013, lin_particle_2013, zeissler_diameter-independent_2020,fert_magnetic_2017}.
The existence of a finite
skyrmion Hall angle is an issue for technological applications,
since it can cause skyrmions to move towards the
sample edge and be annihilated,
leading to data loss in the skyrmion-based device.
In order to prevent this, there have been extensive studies on
ways to precisely control the skyrmion motion.
Methods proposed to
mitigate the skyrmion Hall angle, or to
take advantage of some of its proprieties,
include
periodic pinning \cite{reichhardt_quantized_2015, reichhardt_nonequilibrium_2018, feilhauer_controlled_2020, vizarim_directional_2021, vizarim_skyrmion_2020, reichhardt_commensuration_2022},
sample curvature \cite{carvalho-santos_skyrmion_2021, korniienko_effect_2020, yershov_curvature-induced_2022},
interface guided motion \cite{vizarim_guided_2021, zhang_edge-guided_2022},
ratchet effects \cite{reichhardt_magnus-induced_2015, souza_skyrmion_2021, chen_skyrmion_2019, gobel_skyrmion_2021, souza_controlled_2024},
temperature and magnetic field gradients \cite{yanes_skyrmion_2019, zhang_manipulation_2018, everschor_rotating_2012, kong_dynamics_2013},
granular films \cite{gong_current-driven_2020,del-valle_defect_2022},
parametric pumping \cite{yuan_wiggling_2019},
voltage-controlled PMA \cite{zhang_magnetic_2015,zhao_ferromagnetic_2020},
skyrmion-vortex coupling using heterostructures \cite{menezes_manipulation_2019, neto_mesoscale_2022},
skyrmion lattice compression \cite{zhang_structural_2022, bellizotti_souza_spontaneous_2023},
laminar flow of skyrmions \cite{zhang_laminar_2023},
soliton motion along skyrmion chains \cite{vizarim_soliton_2022, souza_soliton_2023},
pulse current modulation \cite{du_steady_2024},
strain driven motion \cite{liu_strain-induced_2024}, and
motion by interacting with domain walls \cite{xing_skyrmion_2022}.
Recent investigations into the interaction of skyrmions with
both dc and ac driving have aimed at elucidating 
the impact of the Magnus effect on Shapiro steps.
Reichhardt and Reichhardt \cite{reichhardt_shapiro_2015}
demonstrated that the presence of the Magnus force 
caused a shift of the locking steps to higher dc drives.
Additionally, when a
longitudinal dc drive was combined with a transverse ac drive, they 
observed pronounced Shapiro steps
corresponding to intricate 2D periodic orbits that
cannot be replicated using overdamped particles.
In a 
subsequent study \cite{reichhardt_shapiro_2017},
Reichhardt and Reichhardt also identified the occurrence of
Shapiro spikes as well as negative mobility, where
the skyrmions move in the direction opposite to the dc driving
direction.
More recently, Vizarim {\it et al.} \cite{vizarim_shapiro_2020}
found that
significant skyrmion Hall angle overshoots and reversals
can occur under certain conditions. 
Although these observations offer valuable insights into the behavior of skyrmions under 
simultaneous dc and ac driving, it is worth noting that these simulations were conducted using 
a particle-based model \cite{lin_driven_2013}
that does not account for skyrmion annihilation, creation, or deformations such as internal excitations. Consequently, a more 
detailed and realistic study of the structural and dynamic behavior of skyrmions
under ac and dc driving conditions 
is still lacking.

In this work, using atomistic simulations,
we investigate the dynamical behavior of a single
skyrmion
subjected to a combination of external ac and dc driving
while interacting with a quasi-one-dimensional
array of defects that form a stripe pattern
along the $x$ direction, as illustrated
in Fig.~\ref{fig1}.
We find that when the ac drive is applied
in the $y$ direction and the dc drive is applied in the $x$
direction,
velocity steps appear in which the $x$ velocity remains
constant with increasing
dc drive, as in ordinary Shapiro steps,
while the $y$ velocity magnitude increases.
The amplitude of the ac drive determines
the number of Shapiro steps that appear,
with
higher ac amplitudes giving more Shapiro steps.
This behavior changes above a saturation value of the dc drive,
when
the combination of the ac and dc driving
becomes so strong
that no Shapiro steps can form
and the effect of varying the skyrmion size become minimal.
We also investigate the dynamical response when a fixed
dc drive is applied along the $y$ direction and an ac drive
of increasing amplitude is applied
in either the $x$ or $y$ direction.
When both drives are in the $x$ direction, we find
a regime in which both the $x$ and $y$ velocities
exhibit a series of constant velocity steps separated
by velocity dips.
Our results should be useful for spintronics
devices where precise control of the
skyrmion velocity and angle of motion
are required.

\section{Simulation}
\begin{figure}
    \centering
    \includegraphics[width=\columnwidth]{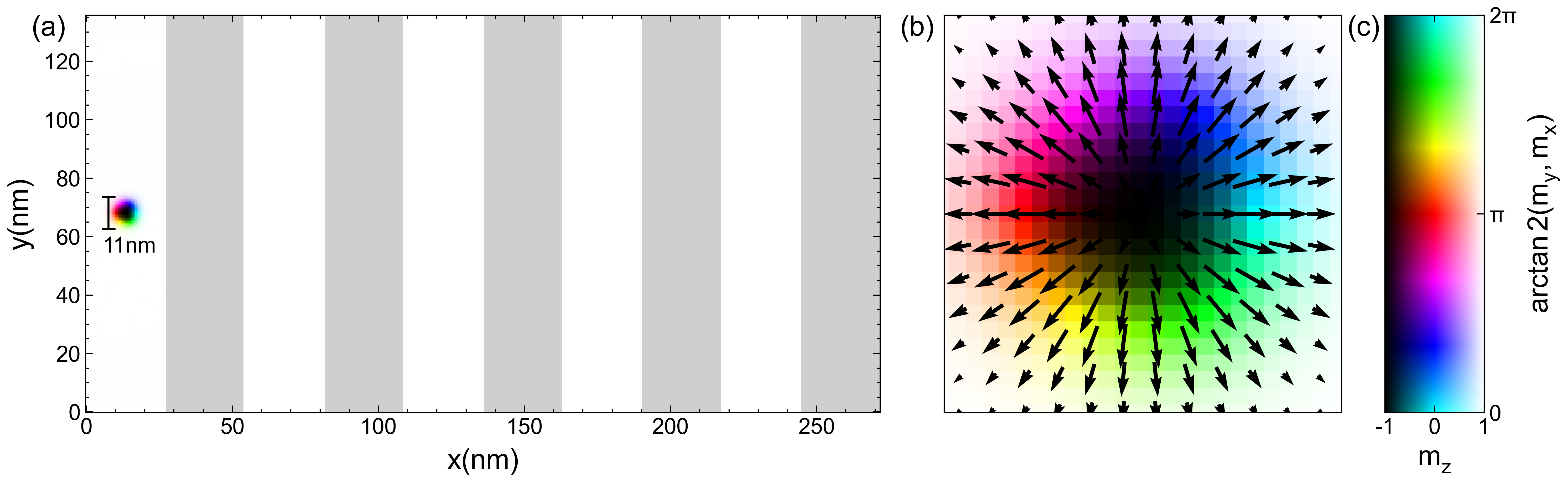}
    \caption{(a) Illustration of the sample used throughout this work.
    Regions with low perpendicular magnetic anisotropy (PMA) of
    $K_L=0.02J$ are colored white
    and regions with high PMA of $K_H=0.05J$ are colored gray.
    We use this anisotropy color representation throughout the work.
    The skyrmion starts at the center left edge of the sample,
    as shown, and has a diameter
    of approximately $\xi=11$ nm for $D/J=0.18$.
    (b) Blow up showing the atomic moments that 
    form the N{\' e}el skyrmion.
    (c) Illustration of the color wheel
    used in this work. Positive out-of-plane magnetic moments
    are white and negative out-of-plane magnetic moments
    are black. Values in between these two limits are colored according
    to the angle between $m_x$ and $m_y$.}
    \label{fig1}
\end{figure}

We consider an ultrathin ferromagnetic
sample that is capable of hosting Néel skyrmions.
The sample has dimensions of 272 nm $\times$ 136 nm and we
apply periodic boundary conditions along the $x$ and $y$ directions.
A magnetic field is applied perpendicular to the sample
surface along the positive $z$ direction at zero temperature,
$T=0$K.
The perpendicular magnetic anisotropy (PMA) takes high and
low values in periodic stripes that extend along the $y$ direction
and vary along the $x$ direction, as illustrated in Fig.~\ref{fig1}(a).
In this work we consider the dynamics of a single
skyrmion under the influence of both dc and ac driving
starting from the initial configuration shown in Fig.~\ref{fig1}(a).

The simulations are performed using the atomistic
model \cite{evans_atomistic_2018}, which captures the dynamics of
individual atomic magnetic moments.
This method enables us to investigate the spin texture
dynamics in detail. The Hamiltonian governing the atomistic
dynamics is given by
\cite{evans_atomistic_2018, iwasaki_universal_2013, iwasaki_current-induced_2013}:

\begin{align}\label{eq1}
  \mathcal{H}=&-\sum_{i, j\in N}J_{i,j}\mathbf{m}_i\cdot\mathbf{m}_j
                -\sum_{i, j\in N}\mathbf{D}_{i,j}\cdot\left(\mathbf{m}_i\times\mathbf{m}_j\right)\\\nonumber
                &-\sum_i\mu\mathbf{H}\cdot\mathbf{m}_i
                -\sum_{i\in R_L} K_L\left(\mathbf{m}_i\cdot\hat{\mathbf{z}}\right)^2\\\nonumber
                &-\sum_{i\in R_H} K_H\left(\mathbf{m}_i\cdot\hat{\mathbf{z}}\right)^2
\end{align}

The underlying lattice is a square arrangement of spins with lattice
constant $a=0.5$ nm.
The first term on the right side is the exchange interaction
between the nearest neighbors contained in the set $N$,
with an exchange constant of $J_{i,j}$ between magnetic moments
$i$ and $j$.
The second term is the interfacial Dzyaloshinskii–Moriya
interaction, where $\mathbf{D}_{i,j}$ is the Dzyaloshinskii–Moriya
vector between magnetic moments $i$ and $j$.
The third term is the Zeeman interaction with an applied external magnetic
field $\mathbf{H}$.
Here $\mu=\hbar\gamma$ is the magnitude of the magnetic moment
and $\gamma=1.76\times10^{11}$T$^{-1}$s$^{-1}$ is the electron
gyromagnetic ratio. The last two terms represent the 
perpendicular magnetic anisotropy (PMA) of the sample.
$K_L$ is the anisotropy constant for low anisotropy
regions contained in the set $R_L$, and $K_H$ is the anisotropy constant
for high anisotropy regions contained in the set $R_H$.
Since we are considering ultrathin films, long-range dipolar interactions
can be neglected because they are expected to be very small
\cite{paul_role_2020}.

The time evolution for the individual
atomic magnetic moments is given by the LLGS
equation \cite{seki_skyrmions_2016, slonczewski_dynamics_1972, gilbert_phenomenological_2004}:

\begin{equation}\label{eq2}
    \frac{\partial\mathbf{m}_i}{\partial t}=-\gamma\mathbf{m}_i\times\mathbf{H}^\text{eff}_i
                             +\alpha\mathbf{m}_i\times\frac{\partial\mathbf{m}_i}{\partial t}
                             +\frac{pa^3}{2e}\left(\mathbf{j}\cdot\nabla\right)\mathbf{m}_i \ .
\end{equation}

Here $\gamma$ is the electron gyromagnetic ratio,
$\mathbf{H}^\text{eff}_i=-\frac{1}{\hbar\gamma}\frac{\partial \mathcal{H}}{\partial \mathbf{m}_i}$
is the effective magnetic field, including all interactions from
the Hamiltonian, $\alpha$ is the phenomenological damping
introduced by Gilbert, and the last term is the adiabatic spin-transfer-torque
(STT), where $p$ is the spin polarization, $e$ the electron
charge, and $\mathbf{j}$ the applied current density.
Use of this STT expression implies that the conduction
electron spins are always parallel to the magnetic moments
$\mathbf{m}$\cite{iwasaki_universal_2013, zang_dynamics_2011}.
The non-adiabatic terms
can be neglected in this case, since they do not affect the skyrmion dynamics
significantly under small driving forces \cite{litzius_skyrmion_2017}.
The current density $\mathbf{j}$ used in this
work has the form:

\begin{equation}\label{eq3}
    \mathbf{j}=j_\text{ac}\left[\cos(2\pi ft)\mathbf{\hat{x}} + \sin(2\pi ft)\mathbf{\hat{y}}\right]
    \cdot\mathbf{\hat{d}}_\text{ac} + j_\text{dc}\mathbf{\hat{d}}_\text{dc}
\end{equation}
where the oscillation frequency $f$ is fixed at $f=50.6\times10^6$Hz,
$t$ is the time, and
$\mathbf{\hat{d}}_\text{dc}$ and $\mathbf{\hat{d}}_\text{ac}$
are the directions of the applied dc and ac currents,
respectively.

The skyrmion velocity is computed using the emergent
electromagnetic fields \cite{seki_skyrmions_2016, schulz_emergent_2012}:

\begin{eqnarray}\label{eq4}
    E^\text{em}_i=\frac{\hbar}{e}\mathbf{m}\cdot\left(\frac{\partial \mathbf{m}}{\partial i}\times\frac{\partial \mathbf{m}}{\partial t}\right)\\
    B^\text{em}_i=\frac{\hbar}{2e}\varepsilon_{ijk}\mathbf{m}\cdot\left(\frac{\partial \mathbf{m}}{\partial j}\times\frac{\partial \mathbf{m}}{\partial k}\right)
\end{eqnarray}
where $\varepsilon_{ijk}$ is the totally anti-symmetric tensor.
The skyrmion drift velocity, $\mathbf{v}_d$, is then calculated using $\mathbf{E}^\text{em}=-\mathbf{v}_d\times\mathbf{B}^\text{em}$.
Our simulation uses fixed values of $\mu\mathbf{H}=0.5(D^2/J)\mathbf{\hat{z}}$,
$\alpha=0.3$, and $p=-1.0$.
The material parameters, unless otherwise specified, are $J=1$meV,
$D=0.18J$, $K_L=0.02J$ and $K_H=0.05J$. These material parameters
stabilize Néel skyrmions similar to those found for Pt/Co/MgO thin-films \cite{boulle_room-temperature_2016}.
For each simulation, the system starts in the initial
configuration illustrated in Fig.~\ref{fig1}(a).
The numerical integration of Eq.~\ref{eq2} is performed using
a fourth order Runge-Kutta method.
For each value of $j_\text{ac}$ and $j_\text{dc}$,
we calculate the time average skyrmion velocity
components, $\left\langle v_x\right\rangle$ and $\left\langle v_y\right\rangle$,
over $3\times10^{7}$ timesteps 
to ensure a steady state.

\section{Dc drive along $x$ and ac drive along $y$}
We begin our analysis with a system in which
the ac drive is applied along
the $y$ direction and the dc drive is applied along
the $x$ direction, giving
$\mathbf{\hat{d}}_\text{ac}=\mathbf{\hat{y}}$ and
$\mathbf{\hat{d}}_\text{dc}=\mathbf{\hat{x}}$.
We fix $j_\text{ac}=5\times10^{10}$Am$^{-2}$ and
vary $j_\text{dc}$ over the range
$1\times10^{10}$Am$^{-2}\leq j_\text{dc}\leq 8\times10^{10}$Am$^{-2}$.

\begin{figure}
    \centering
    \includegraphics[width=\columnwidth]{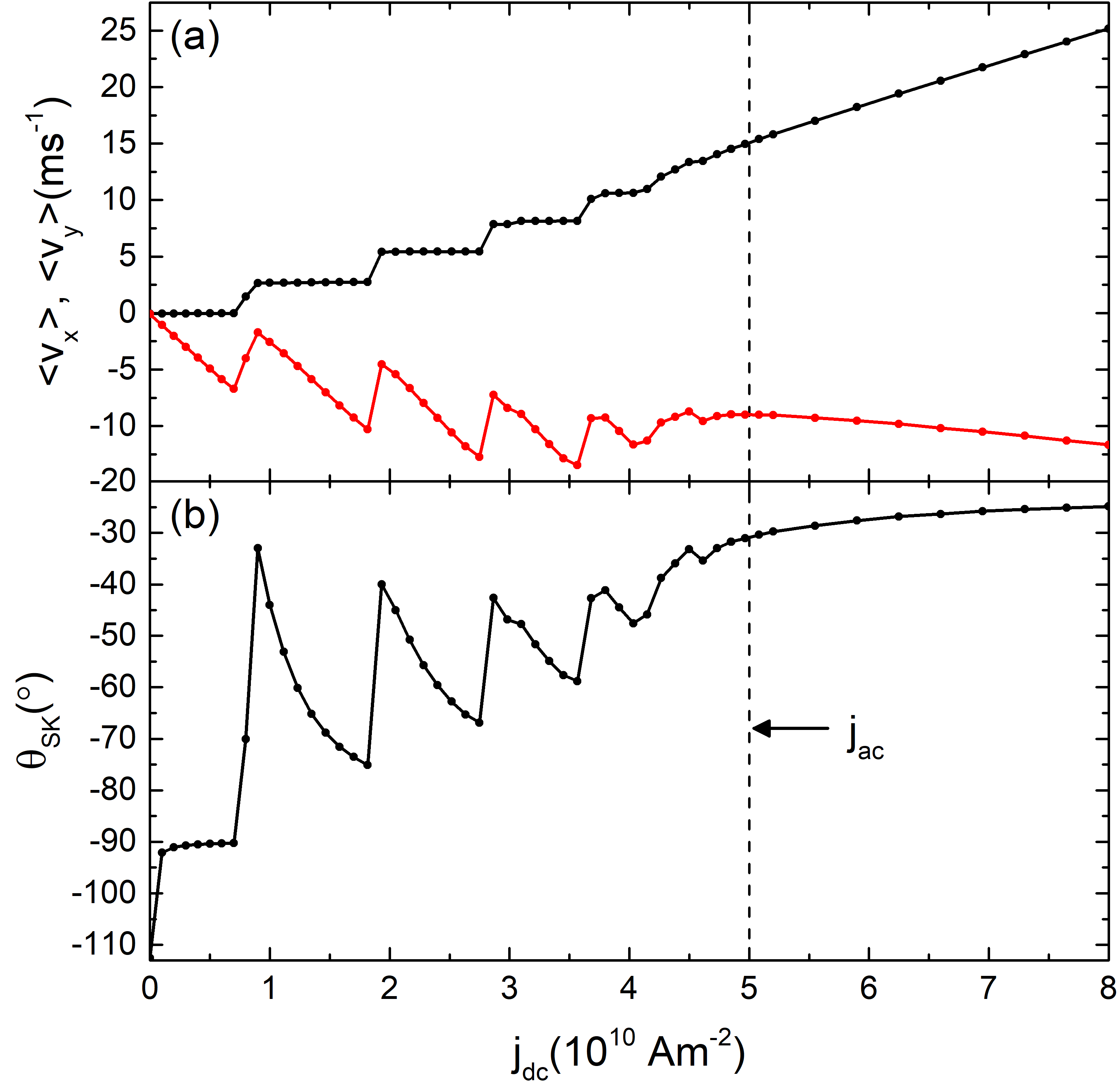}
    \caption{
      A sample with $y$ direction ac driving and $x$ direction dc driving
      where $j_{ac}=5 \times 10^{10}$ Am$^{-2}$ and $D/J=0.18$.
      (a) The average skyrmion velocities
      $\left\langle v_x\right\rangle$ (black) and
      $\left\langle v_y\right\rangle$ (red) vs $j_{dc}$.
      (b) The corresponding skyrmion Hall angle $\theta_{sk}$ vs $j_{dc}$.
      The vertical dashed line indicates the value of $j_{dc}$ at which
      $j_{dc}=j_{ac}$.
    }
    \label{fig2}
\end{figure}

In Fig.~\ref{fig2} we plot
the average velocities $\left\langle v_x\right\rangle$ and
$\left\langle v_y\right\rangle$ along with the skyrmion Hall angle
$\theta_{sk}=\arctan\left(\left\langle v_y\right\rangle/\left\langle v_x\right\rangle\right)$ as 
a function of $j_\text{dc}$.
Here, $\left\langle v_x\right\rangle$ exhibits
constant velocity steps similar to the Shapiro steps that appear when the
ac and dc drives are both applied in the same direction.
In overdamped systems, when the ac and dc drives are
both aligned along the $x$ direction,
velocity steps appear in the $x$ direction but
there are no steps in the $y$ direction since there is no net drift
along $y$; however, the skyrmions have a
finite velocity response in both the $x$ and $y$ directions
due to the gyrotropic force.
The steps in $\langle v_x\rangle$ in Fig.~\ref{fig2}(a)
span a wide range of $j_\text{dc}$, and over this same range, 
Fig.~\ref{fig2}(b) indicates that the skyrmion Hall angle
has a non-monotonic behavior.
Along a given constant
$\langle v_x\rangle$ step, such as the step
appearing for
$1.93\times10^{10}$Am$^{-2}<j_\text{dc}<2.75\times10^{10}$Am$^{-2}$,
the magnitude of 
$\left\langle v_y\right\rangle$ increases linearly with
increasing $j_\text{dc}$.
As a consequence, $\theta_{sk}$ becomes significantly more negative,
meaning that the direction of the skyrmion flow
changes as $j_\text{dc}$ increases along the velocity step.
When a new velocity step is reached and the value
of $\left\langle v_x\right\rangle$ changes,
there is a pronounced jump in $\left\langle v_y\right\rangle$ to
a value that is smaller in magnitude.
As $j_\text{dc}$ increases along this new step,
the magnitude of $\left\langle v_y\right\rangle$ increases linearly again
until the next step is reached and the process repeats.
The velocity steps are associated with
Shapiro step phase locking, where
there is a matching of the ac drive frequency or its higher harmonics
to the frequency of the skyrmion
velocity oscillations produced
by the periodic encounters with the higher anisotropy regions, $R_H$.
These regions
are spaced periodically across the sample and act
as slightly repulsive barriers to the skyrmion motion.
Shapiro steps have been observed previously for skyrmions 
\cite{reichhardt_shapiro_2017,vizarim_shapiro_2020,reichhardt_shapiro_2015};
however, the present work marks the first observation of skyrmion
Shapiro steps
using a detailed atomistic model, which proves that the skyrmion structure
remains stable across the steps.

\begin{figure}
    \centering
    \includegraphics[width=\columnwidth]{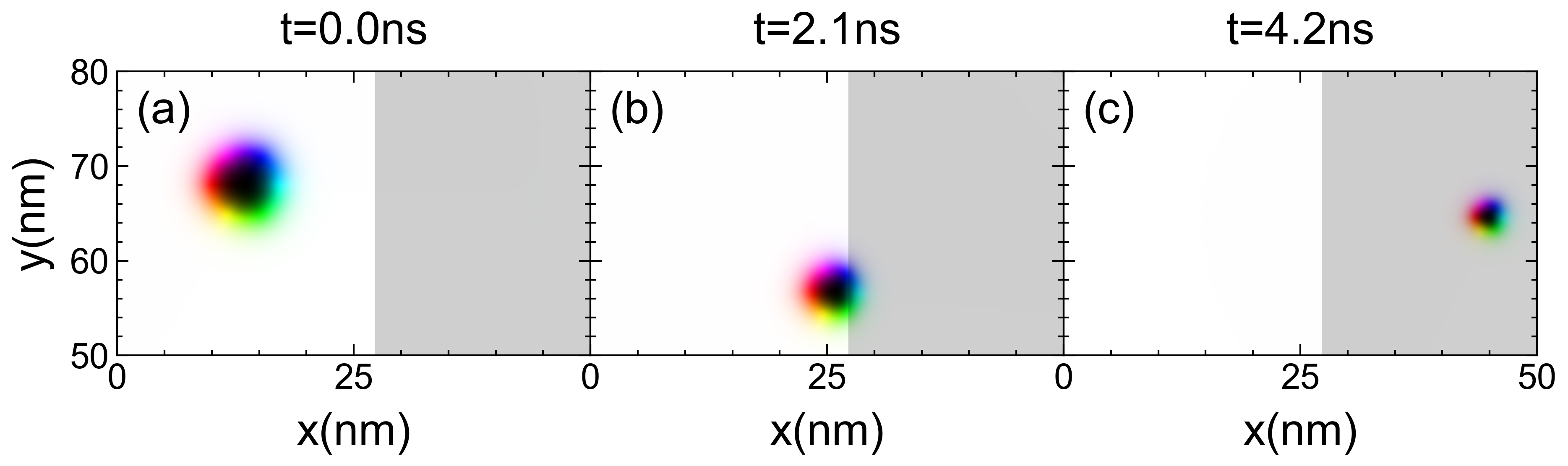}
    \caption{Snapshots of the skyrmion approaching and interacting
    with a region (gray) of stronger perpendicular magnetic anisotropy (PMA)
    for the system from Fig.~\ref{fig2}
    with $y$ direction ac driving and $x$ direction dc driving
    at $j_\text{dc}=2.52\times10^{10}$Am$^{-2}$, 
    $j_\text{ac}=5.00\times10^{10}$Am$^{-2}$, and $D/J=0.18$.
    (a) At $t=0$ ns, the skyrmion is inside the low anisotropy region $R_L$,
    and has its normal diameter of $\xi \approx 11$ nm.
    (b) At $t=2.1$ ns, as the skyrmion approaches the interface
    between regions with different PMA, it shrinks to $\xi \approx 8$ nm.
    (c) At $t=4.2$ ns, when the skyrmion is
    inside region $R_H$ with high anisotropy, $\xi \approx 5$ nm.
    }
    \label{fig3}
\end{figure}

In Fig.~\ref{fig3} we show snapshots of the skyrmion structure
as it passes between
regions with different PMA.
The skyrmion has a normal size
in region $R_L$, but
in region $R_H$ the skyrmion size is strongly reduced. 
During the transition between regions, the skyrmion size changes smoothly.
This size variation as the skyrmion
traverses regions with different PMA is in agreement
with the results of
Refs.~\cite{zhang_magnetic_2015,zhao_ferromagnetic_2020}.
The Shapiro step behavior persists
until $j_\text{dc} \approx j_\text{ac}$, when
both velocity components begin to increase linearly in magnitude
with increasing $j_\text{dc}$
and the step-like behavior of $\left\langle v_x\right\rangle$ is lost.
Note that as $j_\text{dc}$ approaches $j_\text{ac}$, the sharpness of
the Shapiro steps decreases.
The steps are most clearly visible when
$j_\text{dc}$ is small relative to $j_\text{ac}$.
In Fig.~\ref{fig2}(b), the behavior of $\theta_{sk}$ is similar to that
of $\left\langle v_y\right\rangle$.
The increase in magnitude of the skyrmion Hall angle value
along a given step in $\left\langle v_x\right\rangle$ 
is not linear but takes the form
$\theta_{sk}=\arctan\left(\left\langle v_y\right\rangle/\left\langle v_x\right\rangle\right)$ where
$\left\langle v_y\right\rangle$ increases while $\left\langle v_x\right\rangle$ is constant.

\begin{figure}
    \centering
    \includegraphics[width=\columnwidth]{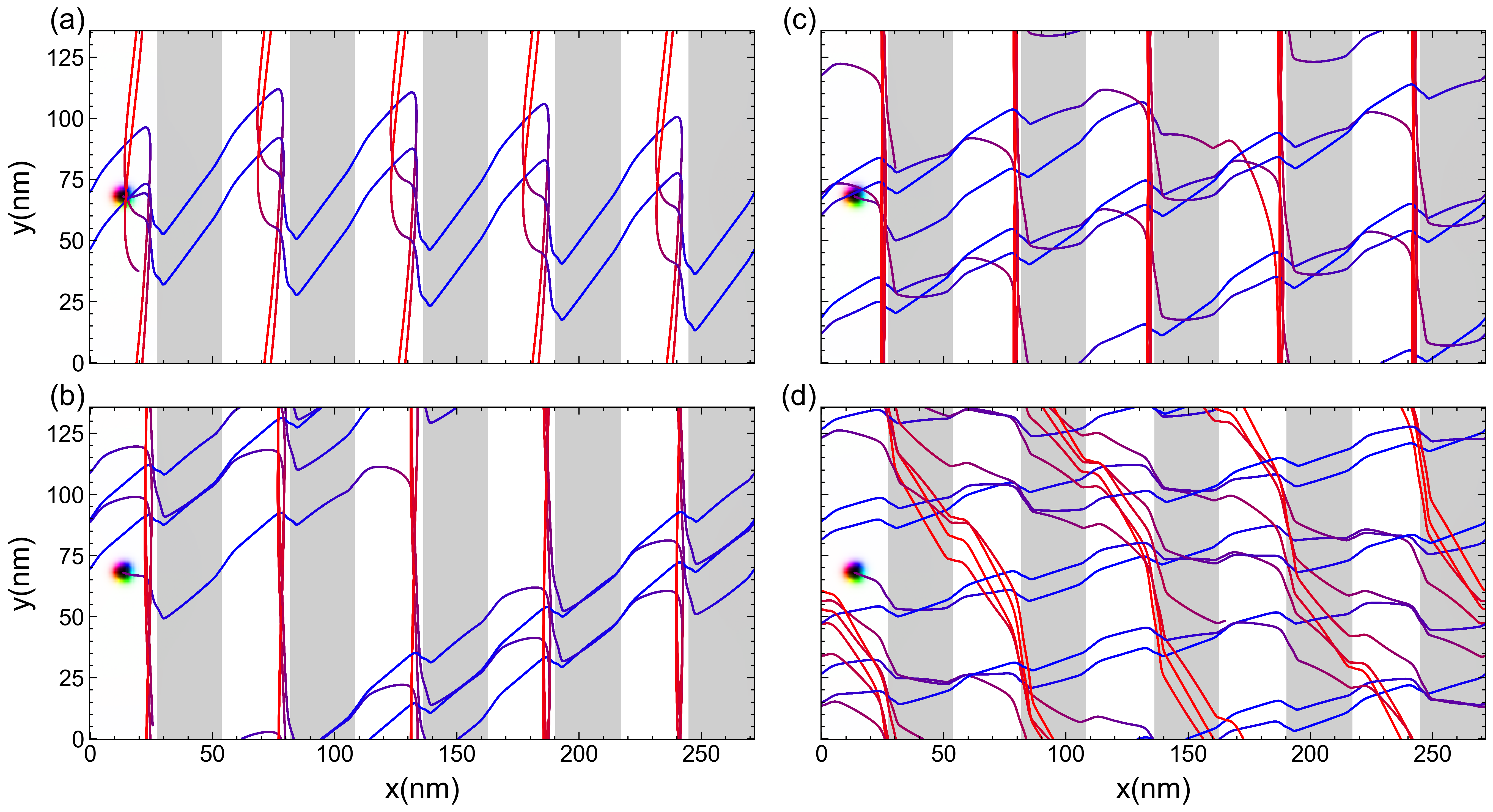}
    \caption{Skyrmion trajectories for the system from Fig.~\ref{fig2}
      with $y$ direction ac driving and $x$ direction dc driving
      at $j_\text{ac}=5.00\times10^{10}$Am$^{-2}$ and $D/J=0.18$.
      The trajectory color varies as a gradient from blue to red
      corresponding to the $+y$ and $-y$ portion of the ac driving
      cycle, respectively.
    White stripes represent lower anisotropy regions and
    gray stripes indicate higher anisotropy regions.
    (a) $j_\text{dc}=1.47\times10^{10}$Am$^{-2}$.
    (b) $j_\text{dc}=2.52\times10^{10}$Am$^{-2}$.
    (c) $j_\text{dc}=3.45\times10^{10}$Am$^{-2}$.
    (d) $j_\text{dc}=5.90\times10^{10}$Am$^{-2}$.}
    \label{fig4}
\end{figure}

In Fig.~\ref{fig4} we illustrate some selected skyrmion trajectories for the
system from Fig.~\ref{fig2}.
At $j_\text{dc}=1.47\times10^{10}$Am$^{-2}$
in Fig.~\ref{fig4}(a),
the $x$ direction skyrmion velocity is constant at
$\left\langle v_x\right\rangle\approx 2.7$ms$^{-1}$.
During the $+y$ portion of the ac drive cycle (blue line), the skyrmion
moves mainly along the $+x$ and $+y$ directions, and
it can traverse exactly two stripes with different PMA values.
In the $-y$ portion of the ac drive cycle (red line),
the skyrmion remains within the low anisotropy region and moves mainly
along the $-y$ direction.
In Fig.~\ref{fig4}(b) at $j_\text{dc}=2.52\times10^{10}$Am$^{-2}$,
$\left\langle v_x\right\rangle\approx 5.4$ms$^{-1}$ and the system is
on the second Shapiro step. Here, during the $+y$ portion of the ac cycle
the skyrmion traverses four stripes,
while the $-y$ motion that appears during the $-y$ portion of the ac cycle
follows nearly straight lines.
For $j_\text{dc}=3.45\times10^{10}$Am$^{-2}$
where $\left\langle v_x\right\rangle\approx 8.1$ms$^{-1}$,
Fig.~\ref{fig4}(c) shows that
the skyrmion
traverses five stripes during the $+y$ portion of the ac cycle.
Finally, when the magnitude of the dc drive is greater
than that of the ac drive, as shown in 
Fig.~\ref{fig4}(d) for $j_\text{dc}=5.90\times10^{10}$Am$^{-2}$,
the skyrmion can traverse multiple stripes
during both the $+y$ and $-y$ portions of the ac drive cycle.
This destroys the resonance between the
ac drive and the collisions between the skyrmion and the $R_H$ regions,
and therefore extinguishes the Shapiro step behavior.

One common dynamical behavior
observed for all values of $j_\text{dc}$ that fall along
a step in $\left\langle v_x\right\rangle$
is that when the skyrmion approaches the
interface between regions $R_L$ and $R_H$, it is repelled
by the higher anisotropy region.
The $R_H$ region thus
acts like a repulsive potential barrier for the skyrmion motion.
The skyrmion can only cross this barrier if it receives enough energy
from the applied drives
to modify its internal structure in order to shrink and
enter the higher anisotropy region, as illustrated in Fig.~\ref{fig3}.
When the ability of the skyrmion to shrink is curtailed by
insufficient available energy from the drive,
so that the skyrmion remains trapped in the lower anisotropy region,
the repulsive interaction causes
the skyrmion to slide close to the $R_L-R_H$ interface,
giving rise to a Magnus velocity boost
of the type observed in previous works
\cite{juge_helium_2021,souza_clogging_2022,bellizotti_souza_spontaneous_2023}.
This behavior can be observed in
Fig.~\ref{fig4}(a,b,c).
During the $+y$ portion of the ac cycle,
the force from the ac drive is rotated by the Magnus term to produce a
skyrmion velocity component along a diagonal from lower left to upper
right. At the same time, the dc driving force that is applied along the $+x$
direction 
generates a Magnus velocity component along a diagonal from upper left
to lower right.
When both of these Magnus components are combined, the net result is
a relatively large $+x$ velocity
that makes it possible for the skyrmion to travel across multiple
stripes with different PMA values.
In contrast, during the $-y$ portion of the ac cycle, the Magnus
force component from the ac drive is oriented along a diagonal from
upper right to lower left. When this motion is combined with the
Magnus term from the dc drive, the $-y$ motion of the skyrmion is enhanced
and the $+x$ motion is diminished, making it
difficult for the skyrmion to cross from one stripe
of PMA to another.

\subsection{Varying the ac amplitude}

\begin{figure}
    \centering
    \includegraphics[width=\columnwidth]{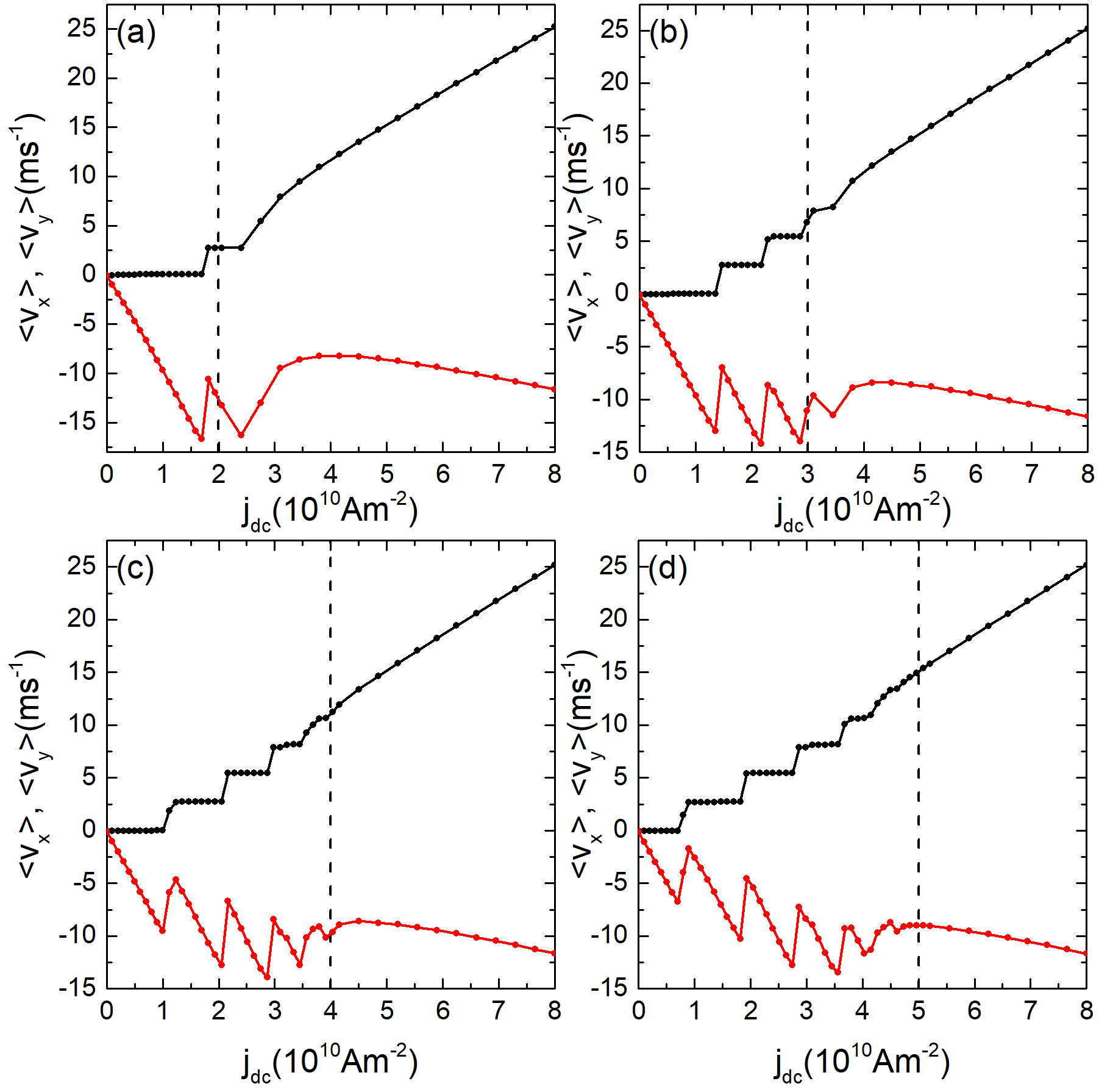}
    \caption{$\left\langle v_x\right\rangle$ (black) and
    $\left\langle v_y\right\rangle$ (red) vs $j_\text{dc}$ in
    samples with $y$ direction ac driving and $x$ direction
    dc driving where $D/J=0.18$.  
    (a) $j_\text{ac}=2\times10^{10}$Am$^{-2}$.
    (b) $j_\text{ac}=3\times10^{10}$Am$^{-2}$.
    (c) $j_\text{ac}=4\times10^{10}$Am$^{-2}$.
    (d) $j_\text{ac}=5\times10^{10}$Am$^{-2}$.
    The vertical dashed lines indicate the value of $j_\text{ac}$ at which
    $j_\text{ac}=j_\text{dc}$.}
    \label{fig5}
\end{figure}

We next explore the impact of the ac drive amplitude on the
dynamical behavior by
considering four different values of $j_\text{ac}$:
$j_\text{ac}=2\times10^{10}\text{Am}^{-2}$,
$3\times10^{10}\text{Am}^{-2}$,
$4\times10^{10}\text{Am}^{-2}$, and
$5\times10^{10}\text{Am}^{-2}$.
In Fig.~\ref{fig5} we plot the skyrmion average velocities
$\left\langle v_x\right\rangle$ and
$\left\langle v_y\right\rangle$ as a function of
the dc drive amplitude $j_\text{dc}$ for each of these ac drive values.
In every case
there is a range of $j_\text{dc}$ over which well defined velocity
Shapiro steps
appear in
$\left\langle v_x\right\rangle$.
On these steps, $\left\langle v_y\right\rangle$ increases linearly in magnitude,
and sudden jumps in $\langle v_y\rangle$ appear
upon transitioning from one $\left\langle v_x\right\rangle$
step to another.
The number of steps that can form is determined
by the ac drive amplitude, with
higher amplitudes favoring the appearance of Shapiro steps and
lower amplitudes hindering step formation.
The range of $j_\text{dc}$ values for which the Shapiro steps
can be observed, which also determines how many
steps are present, is given by
$j_\text{dc}\leq j_\text{ac}$. 
This indicates that Shapiro steps for skyrmion systems should remain
robustly observable for a wide range of ac and dc drive amplitudes.

\subsection{The influence of $D/J$}

\begin{figure}
    \centering
    \includegraphics[width=0.8\columnwidth]{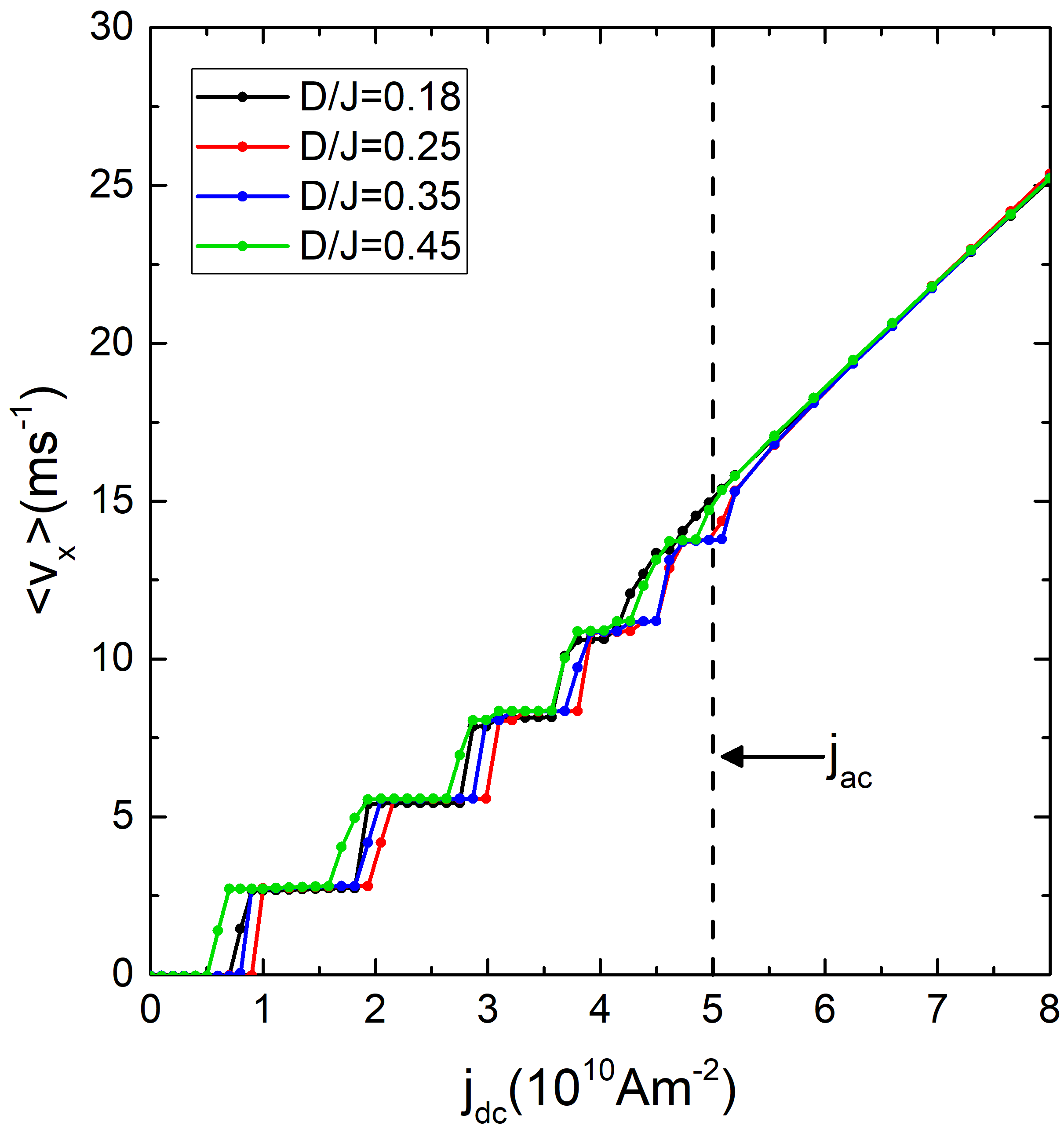}
    \caption{$\left\langle v_x\right\rangle$ vs
    $j_\text{dc}$ for
    $D/J=0.18$ (black),
    $D/J=0.25$ (red),
    $D/J=0.35$ (blue), and
    $D/J=0.45$ (green)
    in samples with $y$ direction ac driving, $x$ direction dc
    driving, and fixed $j_\text{ac}=5\times10^{10}$Am$^{-2}$.
    The dashed line indicates the point at which
    $j_\text{ac}=j_\text{dc}$.
    }
    \label{fig6}
\end{figure}

In skyrmion systems, 
the ratio $D/J$ of magnetic parameters
is responsible for determining the size and stability of
the skyrmion \cite{fert_magnetic_2017}.
Thus, we can investigate the effects of different
skyrmion sizes on the Shapiro steps
by varying the ratio $D/J$.
In Fig.~\ref{fig6} we plot
the average skyrmion velocity
$\left\langle v_x\right\rangle$ as a function of the dc drive intensity
$j_\text{dc}$ for fixed $j_\text{ac}=5\times10^{10}$Am$^{-2}$ and
different values of $D/J$.
We find that Shapiro velocity steps
are present for all values of $D/J$,
indicating that
Shapiro step behavior is robust in this system.
As $D/J$ increases, the Dzyaloshinskii-Moriya interaction becomes stronger than the
exchange interaction, leading to a reduction in the skyrmion size.
The consequences of changing the skyrmion size are subtle,
and we find
that the velocity-current curves can be
shifted towards higher or lower values of $j_\text{dc}$ depending on 
the value of $D/J$.
The number and width of the Shapiro steps 
are very similar for the $D/J$ values simulated here, but we find
that as $D/J$ increases, the transitions between
the Shapiro steps become smoother.
For lower
values of $D/J$, the transitions are sharper and well-defined, whereas
for larger $D/J$ values, transient regions
emerge between the constant velocity steps.
This may be the result of a reduced contrast in the size of
the skyrmion as it moves between the low and high PMA regions when
the skyrmion becomes smaller.
We expect that for very high values
of $D/J$, the transient effects may become dominant
and the Shapiro steps would be lost.

\begin{figure}
    \centering
    \includegraphics[width=\columnwidth]{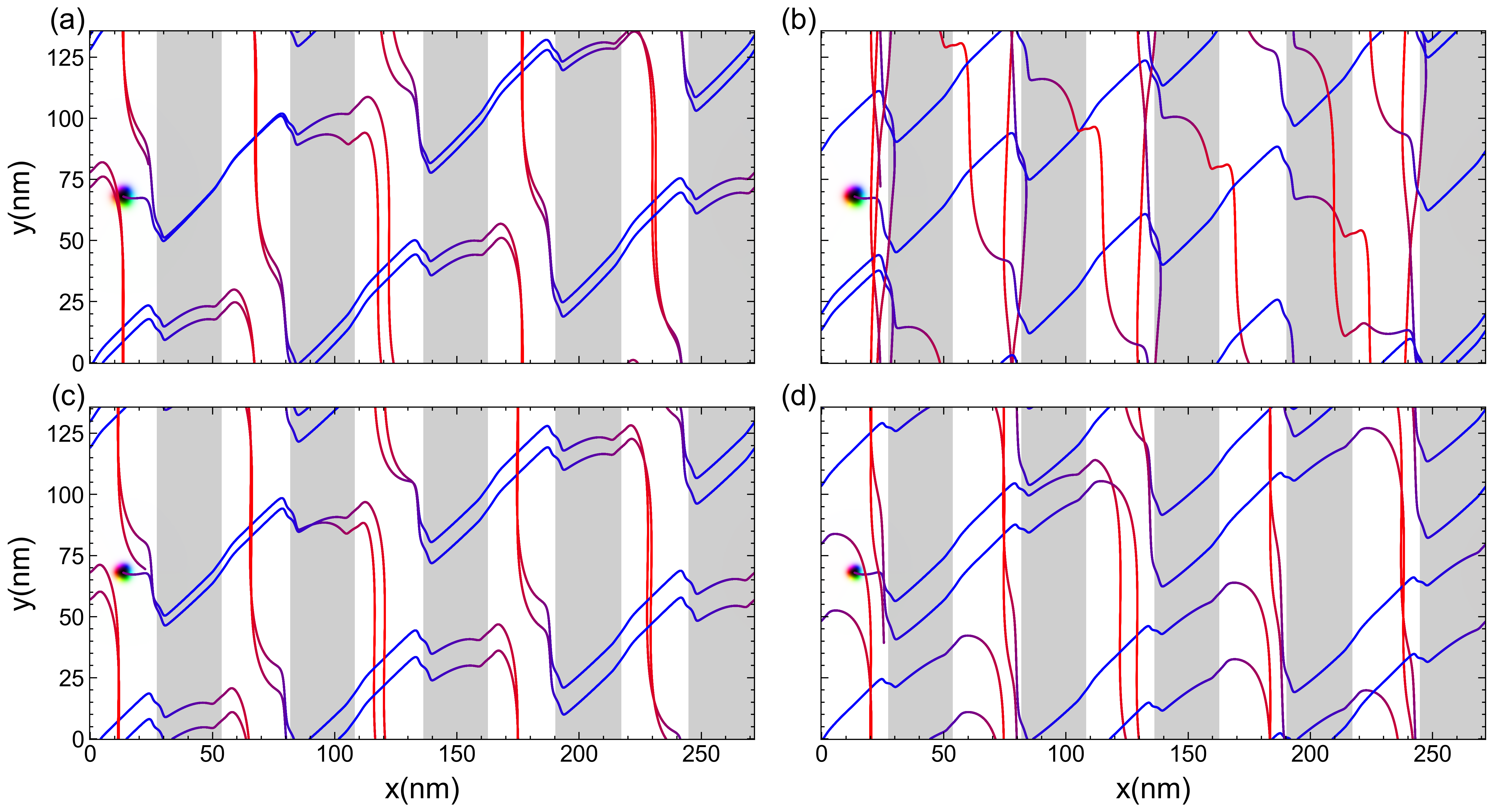}
    \caption{Skyrmion trajectories for the system from Fig.~\ref{fig6}
    with $y$ direction ac driving and $x$ direction dc driving at
    $j_\text{dc}=2.05\times10^{10}$Am$^{-2}$ and
    $j_\text{ac}=5\times10^{10}$Am$^{-2}$. The trajectory color varies
    as a gradient from blue to red corresponding to the $+y$ and $-y$
    portion of the ac driving cycle, respectively. White stripes
    represent lower anisotropy regions and gray stripes indicate
    higher anisotropy regions.
    Note that for each value of $D/J$,
    the skyrmion diameter $\xi$ is slightly different.
    (a) $D/J=0.18$ and $\xi=11.0$ nm.
    (b) $D/J=0.25$ and $\xi=10.8$ nm.
    (c) $D/J=0.35$ and $\xi=9.3$ nm.
    (d) $D/J=0.45$ and $\xi=7.8$ nm.}
    \label{fig7}
\end{figure}

In Fig.~\ref{fig7} we illustrate the skyrmion trajectories
at fixed $j_\text{dc}=2.05\times10^{10}$Am$^{-2}$ and 
$j_\text{ac}=5\times10^{10}$Am$^{-2}$ for different values of $D/J$.
Focusing on the $+y$ portion of the ac cycle, we find that
the skyrmion with $D/J = 0.18$ in Fig.~\ref{fig7}(a) can
traverse approximately two and half stripes during this time period.
In Fig.~\ref{fig7}(b) for $D/J=0.25$, the skyrmion is smaller and it traverses
only two stripes.
For $D/J=0.35$ in Fig.~\ref{fig7}(c), the skyrmion becomes even
smaller but it returns to traveling
approximately two and half stripes.
Finally, for $D/J=0.45$ in Fig.~\ref{fig7}(d), the very small
skyrmion traverses
three stripes.
Although the dynamics change in a non-trivial way,
in each case we find a common behavior: the skyrmion is able to
move from
a low anisotropy region to a high anisotropy region 
only during the $+y$ portion of the ac cycle.
This is the same behavior described in Fig.~\ref{fig4},
where the Magnus force components from the ac and dc drives must
combine in the right way
to allow the skyrmion to overcome the effective potential barrier at the
$R_L-R_H$ interface.

\section{Ac and dc drives in different configurations}

Up until this point,
we have have only considered systems where the
ac drive was applied along the $y$ direction and the dc drive
was applied along the $+x$ direction.
It is, however, known that
other combinations of ac and dc driving can produce interesting
particle dynamics \cite{reichhardt_magnus-induced_2015,reichhardt_shapiro_2015,reichhardt_shapiro_2017,vizarim_shapiro_2020}.
Thus, we next consider the effect of the external drive directions on
the skyrmion dynamics.
We investigate two combinations:
(i) ac drive applied along the $x$ direction
and dc drive applied along the $+y$ direction,
$\mathbf{\hat{d}}_\text{ac}=\mathbf{\hat{x}}$ and
$\mathbf{\hat{d}}_\text{dc}=\mathbf{\hat{y}}$;
and
(ii) ac and dc drives applied along the $x$ direction,
$\mathbf{\hat{d}}_\text{ac}=\mathbf{\hat{d}}_\text{dc}=\mathbf{\hat{x}}$.
In both cases, we fix the
external dc drive at $j_\text{dc}=4\times10^{10}$Am$^{-2}$
and vary the ac drive amplitude $j_\text{ac}$.
Note that this is the opposite to the protocol used in Section III,
where the dc drive was varied while the ac drive amplitude
was held fixed.

\begin{figure}
    \begin{minipage}{0.5\textwidth}
    \includegraphics[width=0.5\columnwidth]{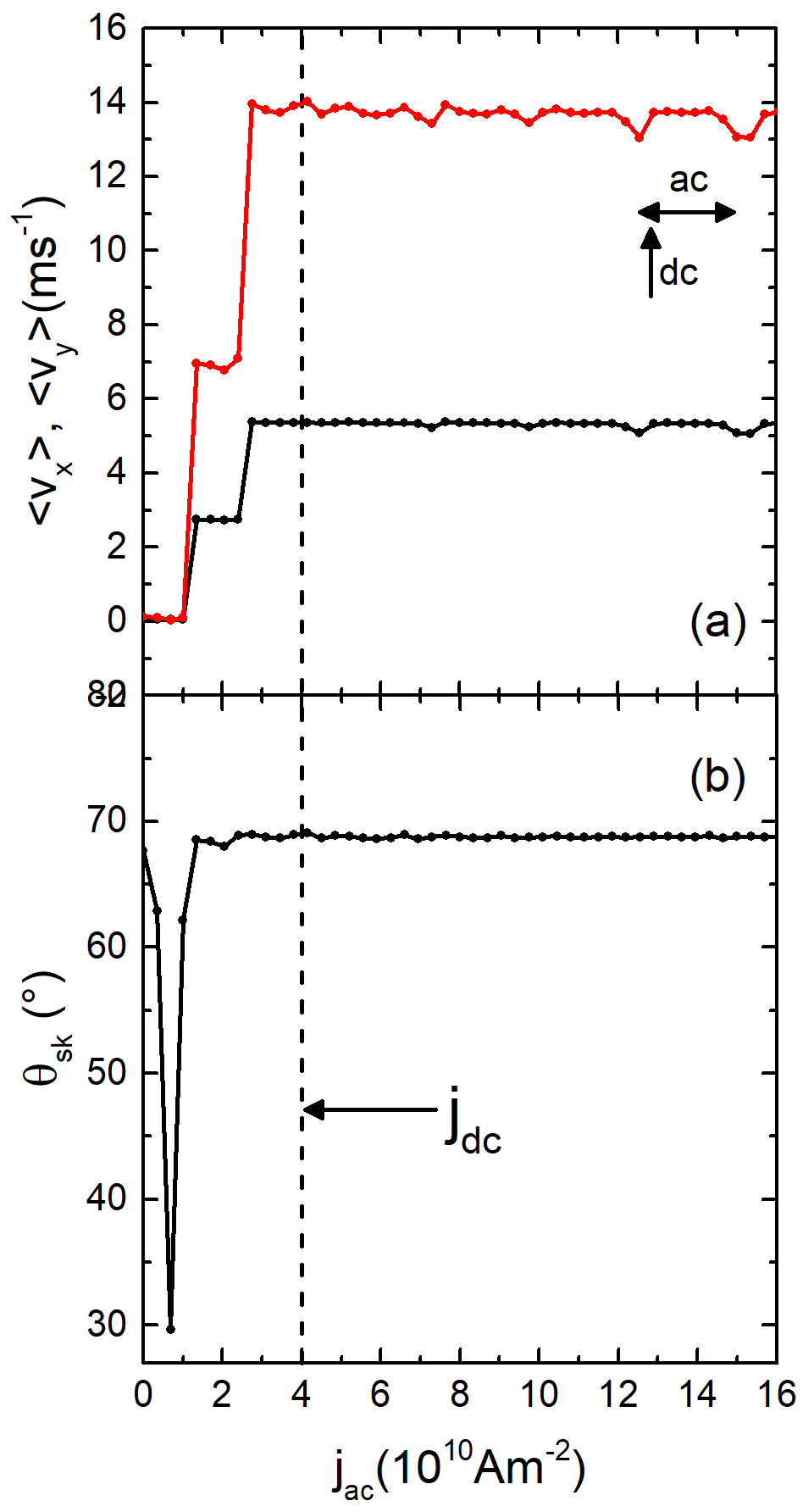}%
    \includegraphics[width=0.5\columnwidth]{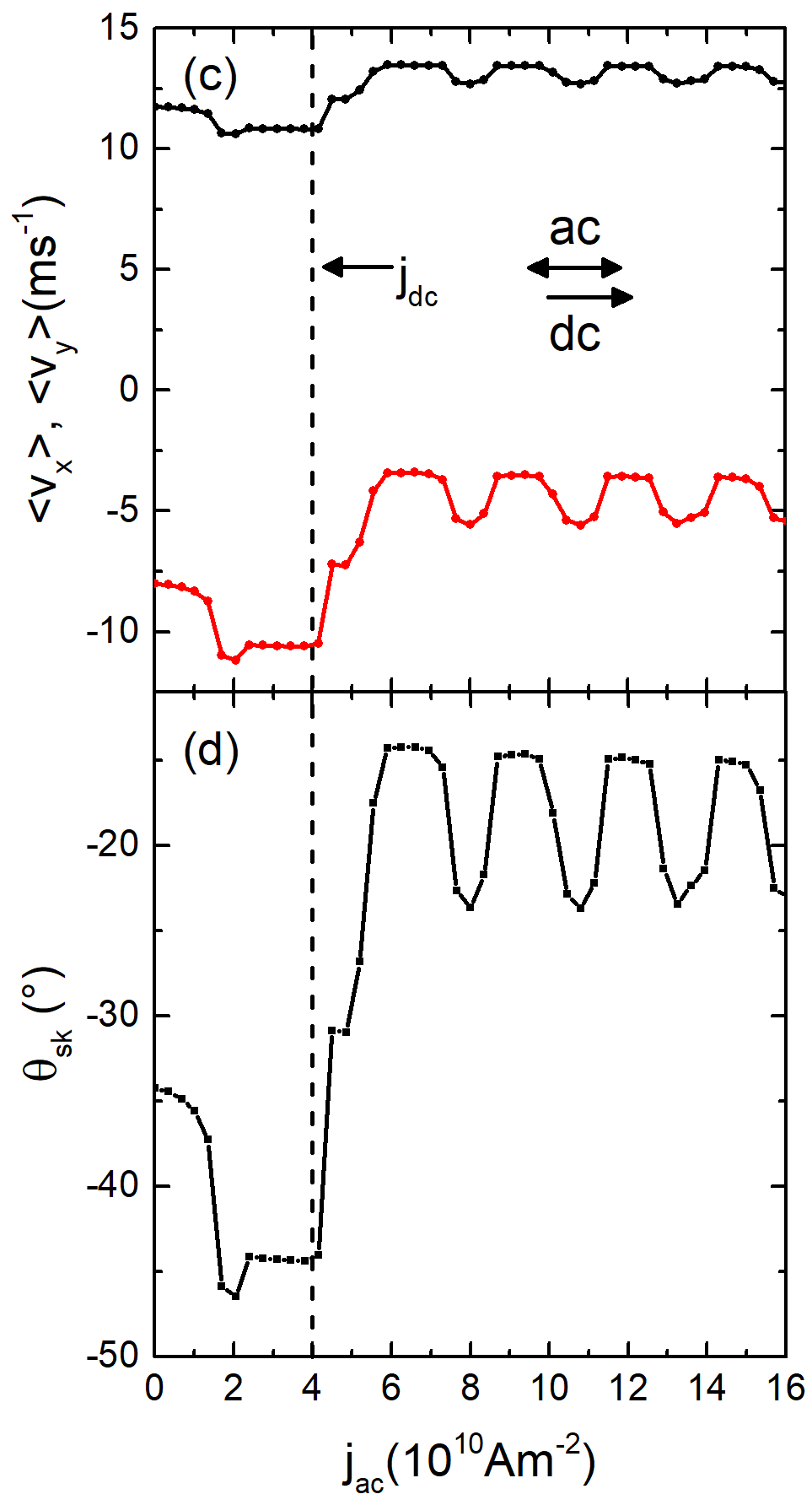}
    \end{minipage}
    \caption{(a,c) $\left\langle v_x\right\rangle$ (black) and
      $\left\langle v_y\right\rangle$ (red) vs $j_\text{ac}$ and
      (b,d) $\theta_{sk}$ vs $j_\text{ac}$
      for samples with fixed $j_\text{dc}=4\times10^{10}$Am$^{-2}$ and
      $D/J=0.18$.
      (a,b) ac drive applied along the $x$ direction and dc
      drive applied along the $+y$ direction.
      (c,d) ac and dc drives applied along the $x$ direction.
      The vertical dashed lines indicate
      the point at which $j_\text{ac}=j_\text{dc}$.
    }
    \label{fig8}
\end{figure}

In Fig.~\ref{fig8}(a) we plot $\left\langle v_x\right\rangle$ and
$\left\langle v_y\right\rangle$ as a function of
$j_\text{ac}$ when the ac drive is applied along the $x$ direction and
the dc drive is
applied along the $+y$ direction.
In the absence of an ac drive, although the $y$ direction dc drive would
produce a component of skyrmion motion along the $x$ direction due to the
Magnus term, this component is not large enough to push the skyrmion across
the barrier at the $R_L-R_H$ interface, and the skyrmion would remain in
the low anisotropy region.
When a finite $x$ direction ac drive is added, however,
the skyrmion is able to traverse the high anisotropy region and
exhibits Shapiro steps with
constant values of 
$\left\langle v_x\right\rangle$.
On these steps, the
ac driving frequency resonates with the rate at which the skyrmion
is crossing the periodic substrate pattern.
When $j_\text{ac}>j_\text{dc}$, the system remains locked on a single
step and there are only small oscillations around the velocities of
$\left\langle v_x\right\rangle \approx 5$ m/s
and $\left\langle v_y\right\rangle \approx 14$ m/s.
This behavior is more like what would be expected for
a Kapitza pendulum \cite{reichhardt_phase-locking_2000,Landau76,Reichhardt02a}, where
the width of the locking steps grow
monotonically with increasing ac drive amplitude
rather than oscillating as in the case of Shapiro steps.

\begin{figure}
    \centering
    \includegraphics[width=\columnwidth]{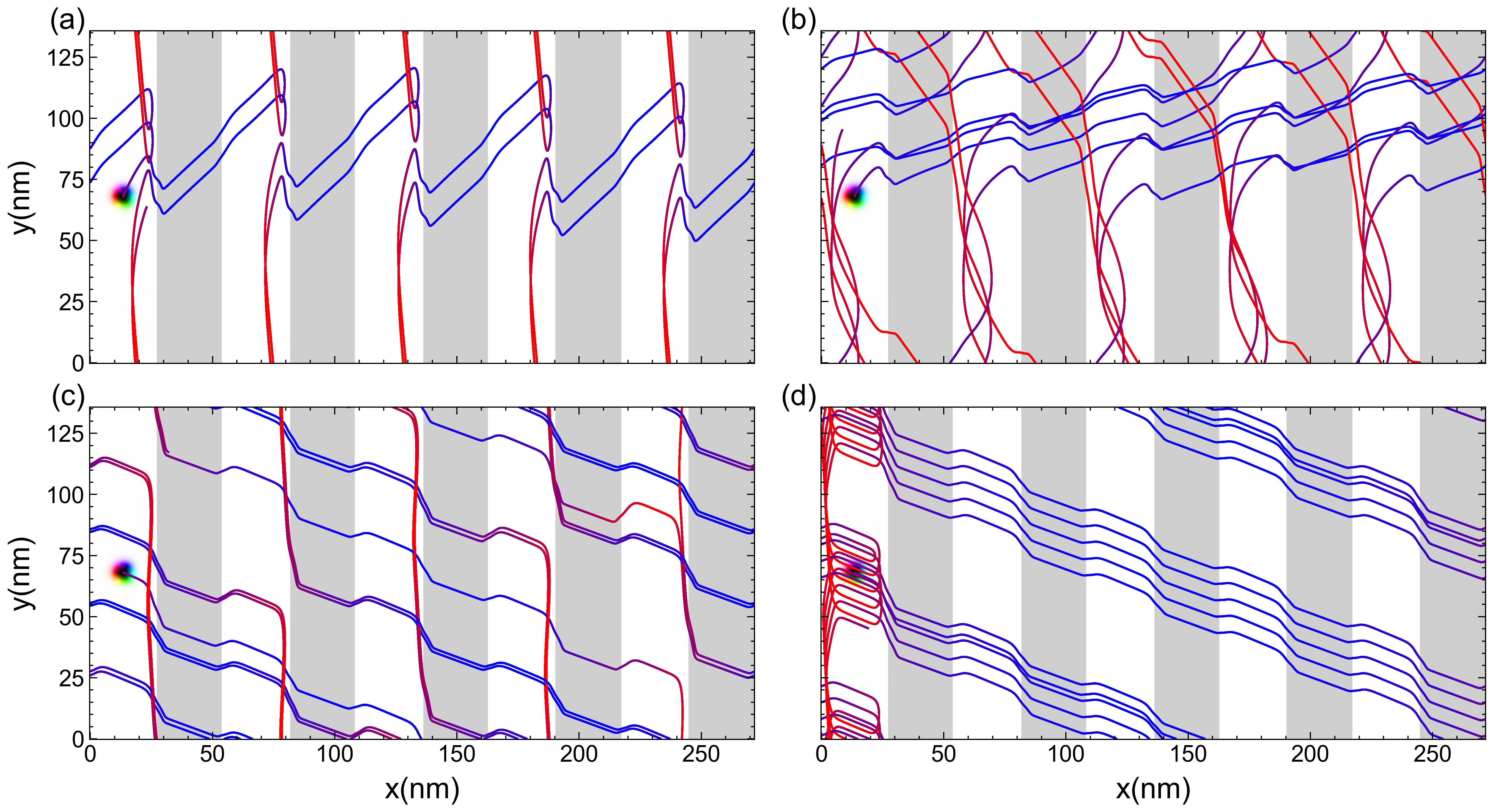}
    \caption{Skyrmion trajectories for systems with
      $j_\text{dc}=4.0\times10^{10}$Am$^{-2}$, $D/J=0.18$,
      and ac driving applied along the $x$ direction.
    The trajectory color varies  
    as a gradient from
    blue to red corresponding to the $+y$ and $-y$ portion of
    the ac driving cycle, respectively. White stripes represent
    lower anisotropy regions and gray stripes indicate higher anisotropy
    regions.
    (a,b) dc driving applied along the $+y$ direction with
    (a) $j_\text{ac}=2.05\times10^{10}$Am$^{-2}$ and
    (b) $j_\text{ac}=5.89\times10^{10}$Am$^{-2}$.
    (c,d) dc driving applied along the $+x$ direction with
    (c) $j_\text{ac}=3.10\times10^{10}$Am$^{-2}$ and
    (d) $j_\text{ac}=6.95\times10^{10}$Am$^{-2}$.}
    \label{fig9}
\end{figure}

In Fig.~\ref{fig9}(a,b) we illustrate the skyrmion trajectories
for the system in Fig.~\ref{fig8}(a) at 
$j_\text{ac}=2.05\times10^{10}$Am$^{-2}$
and $j_\text{ac}=5.89\times10^{10}$Am$^{-2}$, respectively.
The plot of $\theta_{sk}$ versus $j_\text{ac}$ in
Fig.~\ref{fig8}(b) indicates
that the average direction of skyrmion motion
remains
locked to $\theta_{sk} \approx 69^{\circ}$ over the
entire range of $j_\text{ac}$ values considered here.
Similarly to the behavior shown in Section III,
the skyrmion traverses the high anisotropy regions during the positive
portion of the 
ac drive cycle.
In Fig.~\ref{fig9}(a) the skyrmion traverses one stripe
during the $+x$ portion of the ac drive cycle
and is displaced only slightly
along $-x$ during the $-x$ portion of the ac drive cycle
due to the low amplitude of the ac drive.
When the ac drive amplitude is larger, as
in Fig.~\ref{fig9}(b), the skyrmion moves \textit{backwards} by one
stripe during the $-x$ portion of the ac drive cycle.
The interaction of the skyrmion with the barrier at the $R_L-R_H$ interface
changes during each portion of the drive cycle.
For the $+x$ portion of the cycle,
the skyrmion tends to slide along the $-y$ direction, while
during the $-x$ portion of the cycle it slides along
the $+y$ direction.
This behavior enables the skyrmion to jump over the barrier during the
$-x$ portion of the ac drive cycle
and therefore causes
the average skyrmion velocity
to remain constant as long as $j_\text{ac}>j_\text{dc}$,
as shown in Fig.~\ref{fig8}(a).

Figure~\ref{fig8}(c) shows $\langle v_x\rangle$ and $\langle v_y\rangle$
versus $j_\text{ac}$ for samples in which both the ac and dc driving is
applied along the $x$ direction, and
Fig.~\ref{fig8}(d) shows the corresponding
$\theta_{sk}$ versus $j_\text{ac}$ curve.
As illustrated by the trajectories in
Fig.~\ref{fig9}(c,d), for this drive orientation
the skyrmion exhibits a very orderly transport process.
Since both driving currents are applied along the same direction,
the dc drive biases the ac driving cycle and alternately enhances or
reduces the effective ac driving force, making it easier or harder,
respectively, for
the skyrmion to cross the $R_L-R_H$ barriers.
As $j_\text{ac}$ increases
for fixed $j_\text{dc}$,
the velocities pass through a series of
constant velocity Shapiro steps
produced by
resonances between the ac drive frequency and the rate at which the
skyrmion moves across the 
periodic
potential.
Unlike the case for ac driving that is perpendicular to the dc drive,
the skyrmion direction of motion
for parallel drives is not locked in a specific direction,
but changes as a function
of $j_\text{ac}$, as shown in Fig.~\ref{fig8}(d).

The motion of the system in Fig.~\ref{fig8}(c) is
illustrated in Fig.~\ref{fig9}(c) at $j_\text{ac}=3.10\times10^{10}$Am$^{-2}$,
where
the skyrmion follows a zig-zag trajectory.
As the skyrmion crosses
from $R_L$ to $R_H$, it translates slightly along
$-y$, while when it crosses from $R_H$ back to $R_L$,
it moves slightly along $+y$ due to the
interaction with the $R_L-R_H$ interface.
Transport between stripes only occurs during the $+x$ portion of the
ac drive cycle. During the $-x$ portion of the ac drive cycle,
the skyrmion cannot overcome the $R_L-R_H$ barrier
due to the reduction in skyrmion velocity produced by the $+x$ dc drive.
Although the net force is still along $+x$
since $j_\text{dc}>j_\text{ac}$, 
it is not strong enough to push the skyrmion across the potential barrier,
and instead the skyrmion only slides along $-y$.
At $j_\text{ac}=6.95\times10^{10}$Am$^{-2}$ in
Fig.~\ref{fig9}(d),
the net force on the skyrmion is in the $-x$ direction during the $-x$
portion of the ac drive cycle since
$j_\text{ac}>j_\text{dc}$.
Nevertheless,
it is still not large enough to push the skyrmion over
the potential barrier along the $-x$ direction,
resulting in a fast sliding motion of the skyrmion along the $+y$ direction.
The skyrmion can only surpass the potential barrier
during the $+x$ portion of the ac drive cycle,
when both the ac and dc drives combine to
push the skyrmion towards the $+x$ direction with great force,
causing the skyrmion to translate to the right by a distance of nine stripes.

\section{Skyrmion Instability and Future Directions }

\begin{figure}
    \begin{minipage}{0.5\textwidth}
    \includegraphics[width=\columnwidth]{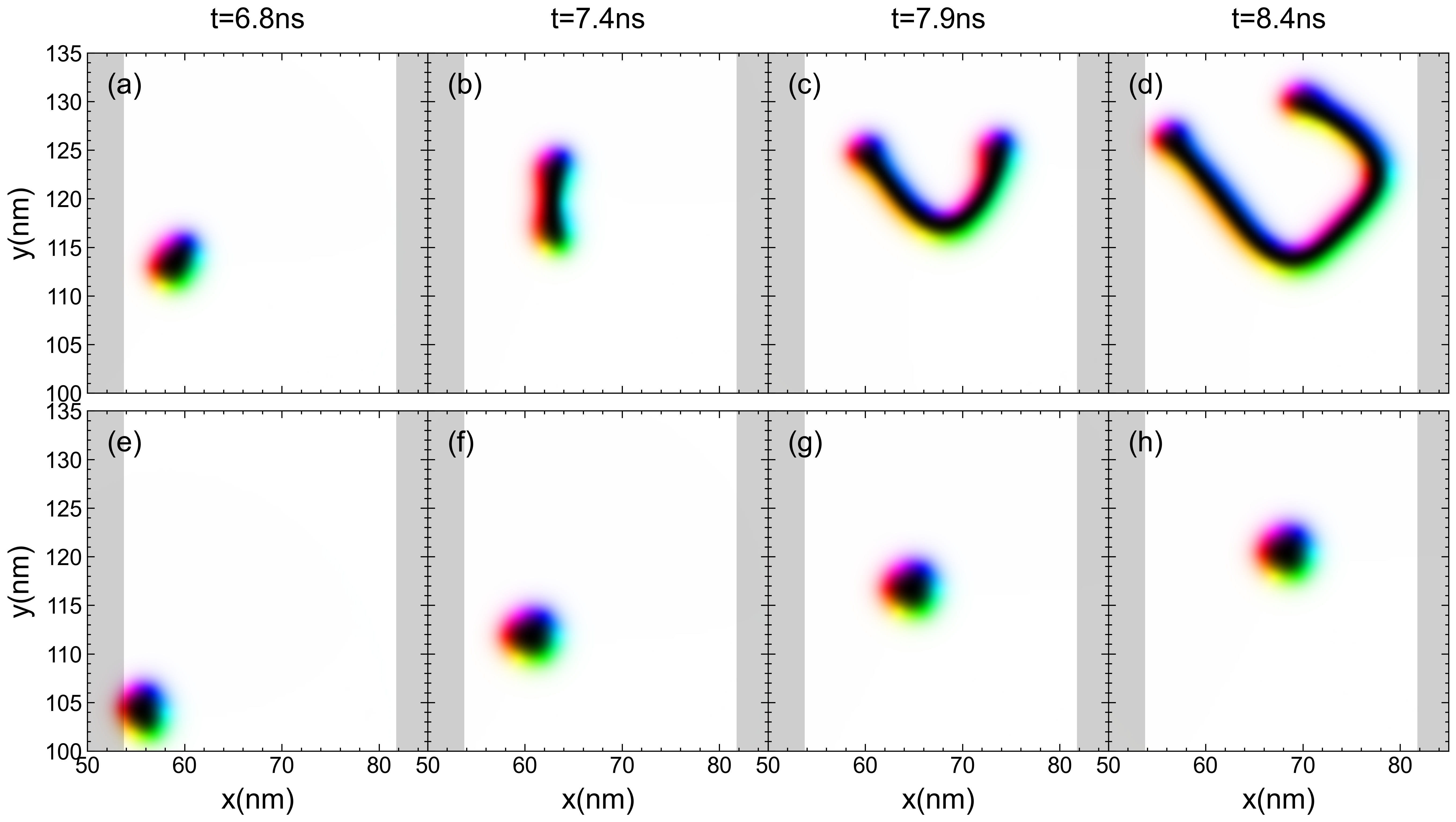}
    \includegraphics[width=\columnwidth]{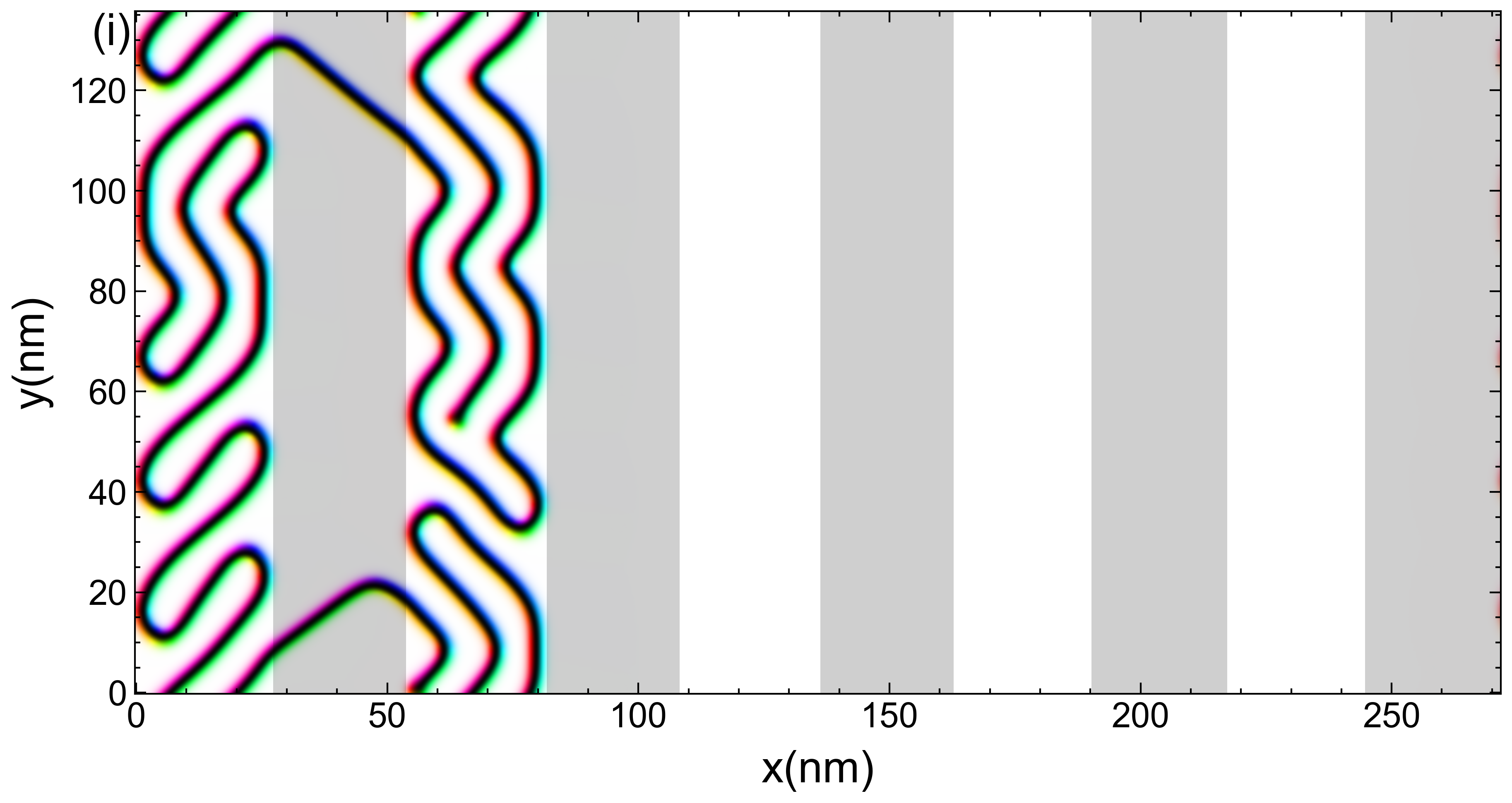}
    \end{minipage}
    \caption{
      Snapshots from systems with $j_\text{dc}=1\times10^{10}$Am$^{-2}$
      and $j_\text{ac}=5\times10^{10}$Am$^{-2}$.
      (a-d) In a system with
      $D/J=0.51$, the skyrmion deforms due to instabilities
      arising during the transition from a high anisotropy region
      (gray) to a low anisotropy region (white).
      (e-h) In a system with $D/J=0.45$, the same transition from
      a high anisotropy region to a low anisotropy region does not cause
      any distortion or instability of the skyrmion.
      (i) The final simulated configuration of the $D/J=0.51$ system
      from panels (a-d)
    showing the appearance of a ``magnetic worm'' after the deformation
    instability. Note that both ends of the worm are in the same low
    anisotropy stripe, with one end in the lower middle of the stripe and
    the other end at the very top of the stripe.
    }
    \label{fig10}
\end{figure}

We find that a range of $D/J$ values exists
in which, instead of well defined skyrmions that produce
Shapiro steps,
an instability arises that transforms
the skyrmion 
into an elongated magnetic texture.
The instability always occurs when the skyrmion is
crossing from the high anisotropy region to the lower anisotropy region, 
$R_H \rightarrow R_L$. It is important to note that in this situation,
the size of the skyrmion is increasing as the crossing occurs.
The skyrmion size is smaller
in higher anisotropy regions and larger in
lower anisotropy regions.
In Fig.~\ref{fig10}(a-d), we illustrate the time evolution of
the transformation of a skyrmion into an
elongated magnetic texture at $D/J=0.51$.
We observe very similar results for the range $0.51<D/J<0.85$.
When $D/J \approx 1$, the skyrmion is usually very small,
and smaller skyrmions tend to be more stable. Additionally,
when the skyrmion radius becomes very small, variations in the skyrmion
radius when passing from one anisotropy region to another also become
quite small, so no instability can arise.
Outside of the instability window, a stable skyrmion can pass
from the high anisotropy to low anisotropy region without distorting
in shape, as illustrated
in Fig.~\ref{fig10}(e-h) for $D/J=0.45$.
When the instability occurs, it can become quite extensive and can
result in an expansion of the skyrmion
into shapes similar to that shown in Fig.~\ref{fig10}(i).
The instability may arise
due to a velocity difference in the skyrmion motion in the two regions
which causes a portion of the skyrmion to expand outward.
This instability will be
addressed in greater detail in
a future work to better understand the topological transition. 
For example, it could also occur for simpler 
sample geometries containing only a single interface between regions
of different anisotropy.
Another question is how important the ac driving is to
producing the instability.

There are many possible future directions for this system,
such as considering multiple skyrmions
or exploring the effect of adding a magnetic field or a finite temperature.
To create a sample of the type we consider in an experiment, it may be
possible to use
periodic thickness modulations
or to pattern the sample with stripes of irradiation.
We focused on a one-dimensional modulation of the anisotropy,
but similar effects could occur for skyrmions moving over
periodic substrates that have two-dimensional variations.
If some asymmetry, such as a sawtooth gradient, were introduced
to the stripes, various ratchet effects could appear.
In addition to skyrmions, it would be interesting to study
whether similar Shapiro step phenomena could arise for antiskyrmions, merons,
and bimerons.
It should also be possible to use the Shapiro step phenomenon
to create new devices that require the skyrmions to be moved at
specific velocities or over specific distances.

\section{Summary}
Using atomistic simulations, we simulate
the dynamical behavior of a single skyrmion
interacting with an array of stripes
of different anisotropy under a combination of external
ac and dc driving.
When a fixed ac drive is applied
along the $y$ direction
and a dc drive of
varied amplitude is applied along the $+x$ direction,
the average velocity passes through a series of
Shapiro steps. On each step,
$\left\langle v_x\right\rangle$ is constant
but $\left\langle v_y\right\rangle$ increases in
magnitude with increasing dc drive amplitude, and there is
a jump down in
$\langle v_y\rangle$ at the transition between adjacent
Shapiro steps.
The skyrmion Hall angle or angle of skyrmion motion changes
across each step, and the skyrmion trajectories on each step
are distinct.
We show that the Shapiro
steps only appear when $j_\text{dc}\leq j_\text{ac}$,
indicating that $j_\text{ac}$
controls the number of Shapiro steps
that can be observed. Increasing $j_\text{ac}$ increases the number
of Shapiro steps that are present until a saturation value
is reached, above which the Shapiro step behavior is lost.
By varying $D/J$, we vary the size of the skyrmion, and find that
this quantity
does not play a major role in determining the appearance of the
Shapiro steps, although it does
change the sharpness of the transition between adjacent Shapiro steps as
well as the shape of the skyrmion trajectories.
For lower values of $D/J$ the transitions are sharp,
but upon increasing $D/J$, the transitions start to become more
continuous.
We expect that for high values of 
$D/J$, such as $D/J=1$, the Shapiro steps will vanish and be
replaced by
an almost linear relationship between the skyrmion velocity
and the magnitude of the dc drive.

We also show that different combinations of dc and ac driving directions
can lead to different skyrmion behavior.
When the dc drive is fixed along
the $+y$ direction and a varying perpendicular ac drive is
applied along the $x$ direction,
the skyrmion direction of motion is locked and $\theta_{sk}$
remains constant, but Shapiro velocity steps appear.
When the ac drive amplitude is higher than the dc drive magnitude,
the system remains locked along a single direction.
For the case of ac and dc drives both applied along the $x$ direction,
we find oscillations in the average skyrmion velocity
associated with positive and negative contributions of the ac
drive cycle to the dc drive.
Our results confirm many of the results previously
observed in particle-based models of skyrmions
and show that Shapiro step phenomena
for various combinations of driving direction should be a robust
feature in skyrmion systems.
The results presented here can be used
to control the motion of skyrmions by taking advantage of the
Shapiro steps and sinusoidal behaviors
observed for different combinations
of external applied drives.

\acknowledgments
This work was supported by the US Department of Energy through the Los Alamos National Laboratory. Los
Alamos National Laboratory is operated by Triad National Security, LLC, for the National Nuclear Security
Administration of the U. S. Department of Energy (Contract No. 892333218NCA000001). 
J.C.B.S acknowledges funding from Fundação de Amparo à Pesquisa do Estado de São Paulo - FAPESP (Grant 2022/14053-8).
We would like to thank Dr. Felipe F. Fanchini for providing the computational resources used in this work. 
These resources were funded by the Fundação de Amparo à Pesquisa do Estado de São Paulo - FAPESP (Grant: 2021/04655-8).

\bibliography{mybib}

\begin{thebibliography}{88}%
\makeatletter
\providecommand \@ifxundefined [1]{%
 \@ifx{#1\undefined}
}%
\providecommand \@ifnum [1]{%
 \ifnum #1\expandafter \@firstoftwo
 \else \expandafter \@secondoftwo
 \fi
}%
\providecommand \@ifx [1]{%
 \ifx #1\expandafter \@firstoftwo
 \else \expandafter \@secondoftwo
 \fi
}%
\providecommand \natexlab [1]{#1}%
\providecommand \enquote  [1]{``#1''}%
\providecommand \bibnamefont  [1]{#1}%
\providecommand \bibfnamefont [1]{#1}%
\providecommand \citenamefont [1]{#1}%
\providecommand \href@noop [0]{\@secondoftwo}%
\providecommand \href [0]{\begingroup \@sanitize@url \@href}%
\providecommand \@href[1]{\@@startlink{#1}\@@href}%
\providecommand \@@href[1]{\endgroup#1\@@endlink}%
\providecommand \@sanitize@url [0]{\catcode `\\12\catcode `\$12\catcode `\&12\catcode `\#12\catcode `\^12\catcode `\_12\catcode `\%12\relax}%
\providecommand \@@startlink[1]{}%
\providecommand \@@endlink[0]{}%
\providecommand \url  [0]{\begingroup\@sanitize@url \@url }%
\providecommand \@url [1]{\endgroup\@href {#1}{\urlprefix }}%
\providecommand \urlprefix  [0]{URL }%
\providecommand \Eprint [0]{\href }%
\providecommand \doibase [0]{http://dx.doi.org/}%
\providecommand \selectlanguage [0]{\@gobble}%
\providecommand \bibinfo  [0]{\@secondoftwo}%
\providecommand \bibfield  [0]{\@secondoftwo}%
\providecommand \translation [1]{[#1]}%
\providecommand \BibitemOpen [0]{}%
\providecommand \bibitemStop [0]{}%
\providecommand \bibitemNoStop [0]{.\EOS\space}%
\providecommand \EOS [0]{\spacefactor3000\relax}%
\providecommand \BibitemShut  [1]{\csname bibitem#1\endcsname}%
\let\auto@bib@innerbib\@empty
\bibitem [{\citenamefont {Pikovsky}\ \emph {et~al.}(2003)\citenamefont {Pikovsky}, \citenamefont {Rosenblum},\ and\ \citenamefont {Kurths}}]{pikovsky_synchronization_2003}%
  \BibitemOpen
  \bibfield  {author} {\bibinfo {author} {\bibfnamefont {A.}~\bibnamefont {Pikovsky}}, \bibinfo {author} {\bibfnamefont {M.}~\bibnamefont {Rosenblum}}, \ and\ \bibinfo {author} {\bibfnamefont {J.}~\bibnamefont {Kurths}},\ }\href@noop {} {\emph {\bibinfo {title} {Synchronization: A Universal Concept in Nonlinear Sciences: 12}}}\ (\bibinfo  {publisher} {Cambridge University Press},\ \bibinfo {year} {2003})\BibitemShut {NoStop}%
\bibitem [{\citenamefont {Ott}(2002)}]{ott_chaos_2002}%
  \BibitemOpen
  \bibfield  {author} {\bibinfo {author} {\bibfnamefont {E.}~\bibnamefont {Ott}},\ }\href@noop {} {\emph {\bibinfo {title} {Chaos in Dynamical Systems}}},\ \bibinfo {edition} {2nd}\ ed.\ (\bibinfo  {publisher} {Cambridge University Press},\ \bibinfo {year} {2002})\BibitemShut {NoStop}%
\bibitem [{\citenamefont {Bennett}\ \emph {et~al.}(2002)\citenamefont {Bennett}, \citenamefont {Schatz}, \citenamefont {Rockwood},\ and\ \citenamefont {Wiesenfeld}}]{bennett_huygenss_2002}%
  \BibitemOpen
  \bibfield  {author} {\bibinfo {author} {\bibfnamefont {M.}~\bibnamefont {Bennett}}, \bibinfo {author} {\bibfnamefont {M.~F.}\ \bibnamefont {Schatz}}, \bibinfo {author} {\bibfnamefont {H.}~\bibnamefont {Rockwood}}, \ and\ \bibinfo {author} {\bibfnamefont {K.}~\bibnamefont {Wiesenfeld}},\ }\bibfield  {title} {\enquote {\bibinfo {title} {Huygens's clocks},}\ }\href {\doibase 10.1098/rspa.2001.0888} {\bibfield  {journal} {\bibinfo  {journal} {Proc. Roy. Soc. London A: Math. Phys. Eng. Sci.}\ }\textbf {\bibinfo {volume} {458}},\ \bibinfo {pages} {563--579} (\bibinfo {year} {2002})}\BibitemShut {NoStop}%
\bibitem [{\citenamefont {Glass}(2001)}]{glass_synchronization_2001}%
  \BibitemOpen
  \bibfield  {author} {\bibinfo {author} {\bibfnamefont {L.}~\bibnamefont {Glass}},\ }\bibfield  {title} {\enquote {\bibinfo {title} {Synchronization and rhythmic processes in physiology},}\ }\href {\doibase 10.1038/35065745} {\bibfield  {journal} {\bibinfo  {journal} {Nature (London)}\ }\textbf {\bibinfo {volume} {410}},\ \bibinfo {pages} {277--284} (\bibinfo {year} {2001})}\BibitemShut {NoStop}%
\bibitem [{\citenamefont {Shapiro}(1963)}]{shapiro_josephson_1963}%
  \BibitemOpen
  \bibfield  {author} {\bibinfo {author} {\bibfnamefont {S.}~\bibnamefont {Shapiro}},\ }\bibfield  {title} {\enquote {\bibinfo {title} {Josephson currents in superconducting tunneling: The effect of microwaves and other observations},}\ }\href {\doibase 10.1103/PhysRevLett.11.80} {\bibfield  {journal} {\bibinfo  {journal} {Phys. Rev. Lett.}\ }\textbf {\bibinfo {volume} {11}},\ \bibinfo {pages} {80--82} (\bibinfo {year} {1963})}\BibitemShut {NoStop}%
\bibitem [{\citenamefont {Barone}\ and\ \citenamefont {Paterno}(1982)}]{barone_physics_1982}%
  \BibitemOpen
  \bibfield  {author} {\bibinfo {author} {\bibfnamefont {A.}~\bibnamefont {Barone}}\ and\ \bibinfo {author} {\bibfnamefont {G.}~\bibnamefont {Paterno}},\ }\href@noop {} {\emph {\bibinfo {title} {Physics and Applications of the {J}osephson Effect}}},\ \bibinfo {edition} {1st}\ ed.\ (\bibinfo  {publisher} {Wiley-{VCH}},\ \bibinfo {year} {1982})\BibitemShut {NoStop}%
\bibitem [{\citenamefont {Benz}\ \emph {et~al.}(1990)\citenamefont {Benz}, \citenamefont {Rzchowski}, \citenamefont {Tinkham},\ and\ \citenamefont {Lobb}}]{benz_fractional_1990}%
  \BibitemOpen
  \bibfield  {author} {\bibinfo {author} {\bibfnamefont {S.~P.}\ \bibnamefont {Benz}}, \bibinfo {author} {\bibfnamefont {M.~S.}\ \bibnamefont {Rzchowski}}, \bibinfo {author} {\bibfnamefont {M.}~\bibnamefont {Tinkham}}, \ and\ \bibinfo {author} {\bibfnamefont {C.~J.}\ \bibnamefont {Lobb}},\ }\bibfield  {title} {\enquote {\bibinfo {title} {Fractional giant {S}hapiro steps and spatially correlated phase motion in {2D J}osephson arrays},}\ }\href {\doibase 10.1103/PhysRevLett.64.693} {\bibfield  {journal} {\bibinfo  {journal} {Phys. Rev. Lett.}\ }\textbf {\bibinfo {volume} {64}},\ \bibinfo {pages} {693--696} (\bibinfo {year} {1990})}\BibitemShut {NoStop}%
\bibitem [{\citenamefont {Coppersmith}\ and\ \citenamefont {Littlewood}(1986)}]{coppersmith_interference_1986}%
  \BibitemOpen
  \bibfield  {author} {\bibinfo {author} {\bibfnamefont {S.~N.}\ \bibnamefont {Coppersmith}}\ and\ \bibinfo {author} {\bibfnamefont {P.~B.}\ \bibnamefont {Littlewood}},\ }\bibfield  {title} {\enquote {\bibinfo {title} {Interference phenomena and mode locking in the model of deformable sliding charge-density waves},}\ }\href {\doibase 10.1103/PhysRevLett.57.1927} {\bibfield  {journal} {\bibinfo  {journal} {Phys. Rev. Lett.}\ }\textbf {\bibinfo {volume} {57}},\ \bibinfo {pages} {1927--1930} (\bibinfo {year} {1986})}\BibitemShut {NoStop}%
\bibitem [{\citenamefont {Gr{\" u}ner}(1988)}]{gruner_dynamics_1988}%
  \BibitemOpen
  \bibfield  {author} {\bibinfo {author} {\bibfnamefont {G.}~\bibnamefont {Gr{\" u}ner}},\ }\bibfield  {title} {\enquote {\bibinfo {title} {The dynamics of charge-density waves},}\ }\href {\doibase 10.1103/RevModPhys.60.1129} {\bibfield  {journal} {\bibinfo  {journal} {Rev. Mod. Phys.}\ }\textbf {\bibinfo {volume} {60}},\ \bibinfo {pages} {1129--1181} (\bibinfo {year} {1988})}\BibitemShut {NoStop}%
\bibitem [{\citenamefont {Brown}\ \emph {et~al.}(1984)\citenamefont {Brown}, \citenamefont {Mozurkewich},\ and\ \citenamefont {Gr{\" u}ner}}]{brown_subharmonic_1984}%
  \BibitemOpen
  \bibfield  {author} {\bibinfo {author} {\bibfnamefont {S.~E.}\ \bibnamefont {Brown}}, \bibinfo {author} {\bibfnamefont {G.}~\bibnamefont {Mozurkewich}}, \ and\ \bibinfo {author} {\bibfnamefont {G.}~\bibnamefont {Gr{\" u}ner}},\ }\bibfield  {title} {\enquote {\bibinfo {title} {Subharmonic {S}hapiro steps and devil's-staircase behavior in driven charge-density-wave systems},}\ }\href {\doibase 10.1103/PhysRevLett.52.2277} {\bibfield  {journal} {\bibinfo  {journal} {Phys. Rev. Lett.}\ }\textbf {\bibinfo {volume} {52}},\ \bibinfo {pages} {2277--2280} (\bibinfo {year} {1984})}\BibitemShut {NoStop}%
\bibitem [{\citenamefont {Martinoli}\ \emph {et~al.}(1975)\citenamefont {Martinoli}, \citenamefont {Daldini}, \citenamefont {Leemann},\ and\ \citenamefont {Stocker}}]{martinoli_c_1975}%
  \BibitemOpen
  \bibfield  {author} {\bibinfo {author} {\bibfnamefont {P.}~\bibnamefont {Martinoli}}, \bibinfo {author} {\bibfnamefont {O.}~\bibnamefont {Daldini}}, \bibinfo {author} {\bibfnamefont {C.}~\bibnamefont {Leemann}}, \ and\ \bibinfo {author} {\bibfnamefont {E.}~\bibnamefont {Stocker}},\ }\bibfield  {title} {\enquote {\bibinfo {title} {{A.C.} quantum interference in superconducting films with periodically modulated thickness},}\ }\href {\doibase 10.1016/0038-1098(75)90043-5} {\bibfield  {journal} {\bibinfo  {journal} {Solid State Commun.}\ }\textbf {\bibinfo {volume} {17}},\ \bibinfo {pages} {205--209} (\bibinfo {year} {1975})}\BibitemShut {NoStop}%
\bibitem [{\citenamefont {Martinoli}(1978)}]{martinoli_static_1978}%
  \BibitemOpen
  \bibfield  {author} {\bibinfo {author} {\bibfnamefont {P.}~\bibnamefont {Martinoli}},\ }\bibfield  {title} {\enquote {\bibinfo {title} {Static and dynamic interaction of superconducting vortices with a periodic pinning potential},}\ }\href {\doibase 10.1103/PhysRevB.17.1175} {\bibfield  {journal} {\bibinfo  {journal} {Phys. Rev. B}\ }\textbf {\bibinfo {volume} {17}},\ \bibinfo {pages} {1175--1194} (\bibinfo {year} {1978})}\BibitemShut {NoStop}%
\bibitem [{\citenamefont {Dobrovolskiy}(2015)}]{dobrovolskiy_ac_2015}%
  \BibitemOpen
  \bibfield  {author} {\bibinfo {author} {\bibfnamefont {O.~V.}\ \bibnamefont {Dobrovolskiy}},\ }\bibfield  {title} {\enquote {\bibinfo {title} {{AC} quantum interference effects in nanopatterned {Nb} microstrips},}\ }\href {\doibase 10.1007/s10948-014-2664-3} {\bibfield  {journal} {\bibinfo  {journal} {J. Supercond. Novel Mag.}\ }\textbf {\bibinfo {volume} {28}},\ \bibinfo {pages} {469--473} (\bibinfo {year} {2015})}\BibitemShut {NoStop}%
\bibitem [{\citenamefont {Van~Look}\ \emph {et~al.}(1999)\citenamefont {Van~Look}, \citenamefont {Rosseel}, \citenamefont {Van~Bael}, \citenamefont {Temst}, \citenamefont {Moshchalkov},\ and\ \citenamefont {Bruynseraede}}]{van_look_shapiro_1999}%
  \BibitemOpen
  \bibfield  {author} {\bibinfo {author} {\bibfnamefont {L.}~\bibnamefont {Van~Look}}, \bibinfo {author} {\bibfnamefont {E.}~\bibnamefont {Rosseel}}, \bibinfo {author} {\bibfnamefont {M.~J.}\ \bibnamefont {Van~Bael}}, \bibinfo {author} {\bibfnamefont {K.}~\bibnamefont {Temst}}, \bibinfo {author} {\bibfnamefont {V.~V.}\ \bibnamefont {Moshchalkov}}, \ and\ \bibinfo {author} {\bibfnamefont {Y.}~\bibnamefont {Bruynseraede}},\ }\bibfield  {title} {\enquote {\bibinfo {title} {Shapiro steps in a superconducting film with an antidot lattice},}\ }\href {\doibase 10.1103/PhysRevB.60.R6998} {\bibfield  {journal} {\bibinfo  {journal} {Phys. Rev. B}\ }\textbf {\bibinfo {volume} {60}},\ \bibinfo {pages} {R6998--R7000} (\bibinfo {year} {1999})}\BibitemShut {NoStop}%
\bibitem [{\citenamefont {Reichhardt}\ \emph {et~al.}(2000)\citenamefont {Reichhardt}, \citenamefont {Scalettar}, \citenamefont {Zim{\' a}nyi},\ and\ \citenamefont {Gr{\o}nbech-Jensen}}]{reichhardt_phase-locking_2000}%
  \BibitemOpen
  \bibfield  {author} {\bibinfo {author} {\bibfnamefont {C.}~\bibnamefont {Reichhardt}}, \bibinfo {author} {\bibfnamefont {R.~T.}\ \bibnamefont {Scalettar}}, \bibinfo {author} {\bibfnamefont {G.~T.}\ \bibnamefont {Zim{\' a}nyi}}, \ and\ \bibinfo {author} {\bibfnamefont {N.}~\bibnamefont {Gr{\o}nbech-Jensen}},\ }\bibfield  {title} {\enquote {\bibinfo {title} {Phase-locking of vortex lattices interacting with periodic pinning},}\ }\href {\doibase 10.1103/PhysRevB.61.R11914} {\bibfield  {journal} {\bibinfo  {journal} {Phys. Rev. B}\ }\textbf {\bibinfo {volume} {61}},\ \bibinfo {pages} {R11914--R11917} (\bibinfo {year} {2000})}\BibitemShut {NoStop}%
\bibitem [{\citenamefont {Sokolovi{\' c}}\ \emph {et~al.}(2017)\citenamefont {Sokolovi{\' c}}, \citenamefont {Mali}, \citenamefont {Odavi{\' c}}, \citenamefont {Rado{\v s}evi{\' c}}, \citenamefont {Medvedeva}, \citenamefont {Botha}, \citenamefont {Shukrinov},\ and\ \citenamefont {Teki{\' c}}}]{sokolovic_devils_2017}%
  \BibitemOpen
  \bibfield  {author} {\bibinfo {author} {\bibfnamefont {I.}~\bibnamefont {Sokolovi{\' c}}}, \bibinfo {author} {\bibfnamefont {P.}~\bibnamefont {Mali}}, \bibinfo {author} {\bibfnamefont {J.}~\bibnamefont {Odavi{\' c}}}, \bibinfo {author} {\bibfnamefont {S.}~\bibnamefont {Rado{\v s}evi{\' c}}}, \bibinfo {author} {\bibfnamefont {S.~Yu.}\ \bibnamefont {Medvedeva}}, \bibinfo {author} {\bibfnamefont {A.~E.}\ \bibnamefont {Botha}}, \bibinfo {author} {\bibfnamefont {Yu.~M.}\ \bibnamefont {Shukrinov}}, \ and\ \bibinfo {author} {\bibfnamefont {J.}~\bibnamefont {Teki{\' c}}},\ }\bibfield  {title} {\enquote {\bibinfo {title} {Devil's staircase and the absence of chaos in the dc- and ac-driven overdamped {Frenkel-Kontorova} model},}\ }\href {\doibase 10.1103/PhysRevE.96.022210} {\bibfield  {journal} {\bibinfo  {journal} {Phys. Rev. E}\ }\textbf {\bibinfo {volume} {96}},\ \bibinfo {pages} {022210} (\bibinfo {year} {2017})}\BibitemShut {NoStop}%
\bibitem [{\citenamefont {Teki{\' c}}\ and\ \citenamefont {Ivi{\' c}}(2011)}]{tekic_frequency_2011}%
  \BibitemOpen
  \bibfield  {author} {\bibinfo {author} {\bibfnamefont {J.}~\bibnamefont {Teki{\' c}}}\ and\ \bibinfo {author} {\bibfnamefont {Z.}~\bibnamefont {Ivi{\' c}}},\ }\bibfield  {title} {\enquote {\bibinfo {title} {Frequency dependence of the subharmonic {Shapiro} steps},}\ }\href {\doibase 10.1103/PhysRevE.83.056604} {\bibfield  {journal} {\bibinfo  {journal} {Phys. Rev. E}\ }\textbf {\bibinfo {volume} {83}},\ \bibinfo {pages} {056604} (\bibinfo {year} {2011})}\BibitemShut {NoStop}%
\bibitem [{\citenamefont {Juniper}\ \emph {et~al.}(2015)\citenamefont {Juniper}, \citenamefont {Straube}, \citenamefont {Besseling}, \citenamefont {Aarts},\ and\ \citenamefont {Dullens}}]{juniper_microscopic_2015}%
  \BibitemOpen
  \bibfield  {author} {\bibinfo {author} {\bibfnamefont {M.~P.~N.}\ \bibnamefont {Juniper}}, \bibinfo {author} {\bibfnamefont {A.~V.}\ \bibnamefont {Straube}}, \bibinfo {author} {\bibfnamefont {R.}~\bibnamefont {Besseling}}, \bibinfo {author} {\bibfnamefont {D.~G. A.~L.}\ \bibnamefont {Aarts}}, \ and\ \bibinfo {author} {\bibfnamefont {R.~P.~A.}\ \bibnamefont {Dullens}},\ }\bibfield  {title} {\enquote {\bibinfo {title} {Microscopic dynamics of synchronization in driven colloids},}\ }\href {\doibase 10.1038/ncomms8187} {\bibfield  {journal} {\bibinfo  {journal} {Nature Commun.}\ }\textbf {\bibinfo {volume} {6}},\ \bibinfo {pages} {7187} (\bibinfo {year} {2015})}\BibitemShut {NoStop}%
\bibitem [{\citenamefont {Brazda}\ \emph {et~al.}(2017)\citenamefont {Brazda}, \citenamefont {July},\ and\ \citenamefont {Bechinger}}]{brazda_experimental_2017}%
  \BibitemOpen
  \bibfield  {author} {\bibinfo {author} {\bibfnamefont {T.}~\bibnamefont {Brazda}}, \bibinfo {author} {\bibfnamefont {C.}~\bibnamefont {July}}, \ and\ \bibinfo {author} {\bibfnamefont {C.}~\bibnamefont {Bechinger}},\ }\bibfield  {title} {\enquote {\bibinfo {title} {Experimental observation of {Shapiro}-steps in colloidal monolayers driven across time-dependent substrate potentials},}\ }\href {\doibase 10.1039/C7SM00393E} {\bibfield  {journal} {\bibinfo  {journal} {Soft Matter}\ }\textbf {\bibinfo {volume} {13}},\ \bibinfo {pages} {4024--4028} (\bibinfo {year} {2017})}\BibitemShut {NoStop}%
\bibitem [{\citenamefont {Abbott}\ \emph {et~al.}(2019)\citenamefont {Abbott}, \citenamefont {Straube}, \citenamefont {Aarts},\ and\ \citenamefont {Dullens}}]{abbott_transport_2019}%
  \BibitemOpen
  \bibfield  {author} {\bibinfo {author} {\bibfnamefont {J.~L.}\ \bibnamefont {Abbott}}, \bibinfo {author} {\bibfnamefont {A.~V.}\ \bibnamefont {Straube}}, \bibinfo {author} {\bibfnamefont {D.~G. A.~L.}\ \bibnamefont {Aarts}}, \ and\ \bibinfo {author} {\bibfnamefont {R.~P.~A.}\ \bibnamefont {Dullens}},\ }\bibfield  {title} {\enquote {\bibinfo {title} {Transport of a colloidal particle driven across a temporally oscillating optical potential energy landscape},}\ }\href {\doibase 10.1088/1367-2630/ab3765} {\bibfield  {journal} {\bibinfo  {journal} {New J. Phys.}\ }\textbf {\bibinfo {volume} {21}},\ \bibinfo {pages} {083027} (\bibinfo {year} {2019})}\BibitemShut {NoStop}%
\bibitem [{\citenamefont {Reichhardt}\ \emph {et~al.}(2001)\citenamefont {Reichhardt}, \citenamefont {Kolton}, \citenamefont {Dom{\' \i}nguez},\ and\ \citenamefont {Gr{\o}nbech-Jensen}}]{reichhardt_phase-locking_2001}%
  \BibitemOpen
  \bibfield  {author} {\bibinfo {author} {\bibfnamefont {C.}~\bibnamefont {Reichhardt}}, \bibinfo {author} {\bibfnamefont {A.~B.}\ \bibnamefont {Kolton}}, \bibinfo {author} {\bibfnamefont {D.}~\bibnamefont {Dom{\' \i}nguez}}, \ and\ \bibinfo {author} {\bibfnamefont {N.}~\bibnamefont {Gr{\o}nbech-Jensen}},\ }\bibfield  {title} {\enquote {\bibinfo {title} {Phase-locking of driven vortex lattices with transverse ac force and periodic pinning},}\ }\href {\doibase 10.1103/PhysRevB.64.134508} {\bibfield  {journal} {\bibinfo  {journal} {Phys. Rev. B}\ }\textbf {\bibinfo {volume} {64}},\ \bibinfo {pages} {134508} (\bibinfo {year} {2001})}\BibitemShut {NoStop}%
\bibitem [{\citenamefont {Marconi}\ \emph {et~al.}(2003)\citenamefont {Marconi}, \citenamefont {Kolton}, \citenamefont {Dom{\' \i}nguez},\ and\ \citenamefont {Gr{\o}nbech-Jensen}}]{marconi_transverse_2003}%
  \BibitemOpen
  \bibfield  {author} {\bibinfo {author} {\bibfnamefont {V.~I.}\ \bibnamefont {Marconi}}, \bibinfo {author} {\bibfnamefont {A.~B.}\ \bibnamefont {Kolton}}, \bibinfo {author} {\bibfnamefont {D.}~\bibnamefont {Dom{\' \i}nguez}}, \ and\ \bibinfo {author} {\bibfnamefont {N.}~\bibnamefont {Gr{\o}nbech-Jensen}},\ }\bibfield  {title} {\enquote {\bibinfo {title} {Transverse phase locking in fully frustrated {Josephson} junction arrays: A different type of fractional giant steps},}\ }\href {\doibase 10.1103/PhysRevB.68.104521} {\bibfield  {journal} {\bibinfo  {journal} {Phys. Rev. B}\ }\textbf {\bibinfo {volume} {68}},\ \bibinfo {pages} {104521} (\bibinfo {year} {2003})}\BibitemShut {NoStop}%
\bibitem [{\citenamefont {Reichhardt}\ \emph {et~al.}(2002)\citenamefont {Reichhardt}, \citenamefont {Olson},\ and\ \citenamefont {Hastings}}]{reichhardt_rectification_2002}%
  \BibitemOpen
  \bibfield  {author} {\bibinfo {author} {\bibfnamefont {C.}~\bibnamefont {Reichhardt}}, \bibinfo {author} {\bibfnamefont {C.~J.}\ \bibnamefont {Olson}}, \ and\ \bibinfo {author} {\bibfnamefont {M.~B.}\ \bibnamefont {Hastings}},\ }\bibfield  {title} {\enquote {\bibinfo {title} {Rectification and phase locking for particles on symmetric two-dimensional periodic substrates},}\ }\href {\doibase 10.1103/PhysRevLett.89.024101} {\bibfield  {journal} {\bibinfo  {journal} {Phys. Rev. Lett.}\ }\textbf {\bibinfo {volume} {89}},\ \bibinfo {pages} {024101} (\bibinfo {year} {2002})}\BibitemShut {NoStop}%
\bibitem [{\citenamefont {Reichhardt}\ and\ \citenamefont {Olson~Reichhardt}(2003)}]{reichhardt_absolute_2003}%
  \BibitemOpen
  \bibfield  {author} {\bibinfo {author} {\bibfnamefont {C.}~\bibnamefont {Reichhardt}}\ and\ \bibinfo {author} {\bibfnamefont {C.~J.}\ \bibnamefont {Olson~Reichhardt}},\ }\bibfield  {title} {\enquote {\bibinfo {title} {Absolute transverse mobility and ratchet effect on periodic two-dimensional symmetric substrates},}\ }\href {\doibase 10.1103/PhysRevE.68.046102} {\bibfield  {journal} {\bibinfo  {journal} {Phys. Rev. E}\ }\textbf {\bibinfo {volume} {68}},\ \bibinfo {pages} {046102} (\bibinfo {year} {2003})}\BibitemShut {NoStop}%
\bibitem [{\citenamefont {Teki{\' c}}\ \emph {et~al.}(2019)\citenamefont {Teki{\' c}}, \citenamefont {Botha}, \citenamefont {Mali},\ and\ \citenamefont {Shukrinov}}]{tekic_inertial_2019}%
  \BibitemOpen
  \bibfield  {author} {\bibinfo {author} {\bibfnamefont {J.}~\bibnamefont {Teki{\' c}}}, \bibinfo {author} {\bibfnamefont {A.~E.}\ \bibnamefont {Botha}}, \bibinfo {author} {\bibfnamefont {P.}~\bibnamefont {Mali}}, \ and\ \bibinfo {author} {\bibfnamefont {Yu.~M.}\ \bibnamefont {Shukrinov}},\ }\bibfield  {title} {\enquote {\bibinfo {title} {Inertial effects in the dc $+$ ac driven underdamped {Frenkel-Kontorova} model: Subharmonic steps, chaos, and hysteresis},}\ }\href {\doibase 10.1103/PhysRevE.99.022206} {\bibfield  {journal} {\bibinfo  {journal} {Phys. Rev. E}\ }\textbf {\bibinfo {volume} {99}},\ \bibinfo {pages} {022206} (\bibinfo {year} {2019})}\BibitemShut {NoStop}%
\bibitem [{\citenamefont {Nagaosa}\ and\ \citenamefont {Tokura}(2013)}]{nagaosa_topological_2013}%
  \BibitemOpen
  \bibfield  {author} {\bibinfo {author} {\bibfnamefont {N.}~\bibnamefont {Nagaosa}}\ and\ \bibinfo {author} {\bibfnamefont {Y.}~\bibnamefont {Tokura}},\ }\bibfield  {title} {\enquote {\bibinfo {title} {Topological properties and dynamics of magnetic skyrmions},}\ }\href {\doibase 10.1038/nnano.2013.243} {\bibfield  {journal} {\bibinfo  {journal} {Nature Nanotechnol.}\ }\textbf {\bibinfo {volume} {8}},\ \bibinfo {pages} {899--911} (\bibinfo {year} {2013})}\BibitemShut {NoStop}%
\bibitem [{\citenamefont {Je}\ \emph {et~al.}(2020)\citenamefont {Je}, \citenamefont {Han}, \citenamefont {Kim}, \citenamefont {Montoya}, \citenamefont {Chao}, \citenamefont {Hong}, \citenamefont {Fullerton}, \citenamefont {Lee}, \citenamefont {Lee}, \citenamefont {Im},\ and\ \citenamefont {Hong}}]{je_direct_2020}%
  \BibitemOpen
  \bibfield  {author} {\bibinfo {author} {\bibfnamefont {S.-G.}\ \bibnamefont {Je}}, \bibinfo {author} {\bibfnamefont {H.-S.}\ \bibnamefont {Han}}, \bibinfo {author} {\bibfnamefont {S.~K.}\ \bibnamefont {Kim}}, \bibinfo {author} {\bibfnamefont {S.~A.}\ \bibnamefont {Montoya}}, \bibinfo {author} {\bibfnamefont {W.}~\bibnamefont {Chao}}, \bibinfo {author} {\bibfnamefont {I.-S.}\ \bibnamefont {Hong}}, \bibinfo {author} {\bibfnamefont {E.~E.}\ \bibnamefont {Fullerton}}, \bibinfo {author} {\bibfnamefont {K.-S.}\ \bibnamefont {Lee}}, \bibinfo {author} {\bibfnamefont {K.-J.}\ \bibnamefont {Lee}}, \bibinfo {author} {\bibfnamefont {M.-Y.}\ \bibnamefont {Im}}, \ and\ \bibinfo {author} {\bibfnamefont {J.-I.}\ \bibnamefont {Hong}},\ }\bibfield  {title} {\enquote {\bibinfo {title} {Direct demonstration of topological stability of magnetic skyrmions \textit{via} topology manipulation},}\ }\href {\doibase 10.1021/acsnano.9b08699} {\bibfield  {journal} {\bibinfo  {journal} {{ACS} Nano}\ }\textbf {\bibinfo {volume} {14}},\
  \bibinfo {pages} {3251--3258} (\bibinfo {year} {2020})}\BibitemShut {NoStop}%
\bibitem [{\citenamefont {Olson~Reichhardt}\ \emph {et~al.}(2014)\citenamefont {Olson~Reichhardt}, \citenamefont {Lin}, \citenamefont {Ray},\ and\ \citenamefont {Reichhardt}}]{olson_reichhardt_comparing_2014}%
  \BibitemOpen
  \bibfield  {author} {\bibinfo {author} {\bibfnamefont {C.~J.}\ \bibnamefont {Olson~Reichhardt}}, \bibinfo {author} {\bibfnamefont {S.~Z.}\ \bibnamefont {Lin}}, \bibinfo {author} {\bibfnamefont {D.}~\bibnamefont {Ray}}, \ and\ \bibinfo {author} {\bibfnamefont {C.}~\bibnamefont {Reichhardt}},\ }\bibfield  {title} {\enquote {\bibinfo {title} {Comparing the dynamics of skyrmions and superconducting vortices},}\ }\href {\doibase 10.1016/j.physc.2014.03.029} {\bibfield  {journal} {\bibinfo  {journal} {Physica C}\ }\textbf {\bibinfo {volume} {503}},\ \bibinfo {pages} {52--57} (\bibinfo {year} {2014})}\BibitemShut {NoStop}%
\bibitem [{\citenamefont {Reichhardt}\ and\ \citenamefont {Reichhardt}(2017{\natexlab{a}})}]{reichhardt_depinning_2016}%
  \BibitemOpen
  \bibfield  {author} {\bibinfo {author} {\bibfnamefont {C.}~\bibnamefont {Reichhardt}}\ and\ \bibinfo {author} {\bibfnamefont {C.~J.~Olson}\ \bibnamefont {Reichhardt}},\ }\bibfield  {title} {\enquote {\bibinfo {title} {Depinning and nonequilibrium dynamic phases of particle assemblies driven over random and ordered substrates: a review},}\ }\href {\doibase 10.1088/1361-6633/80/2/026501} {\bibfield  {journal} {\bibinfo  {journal} {Rep. Prog. Phys.}\ }\textbf {\bibinfo {volume} {80}},\ \bibinfo {pages} {26501} (\bibinfo {year} {2017}{\natexlab{a}})}\BibitemShut {NoStop}%
\bibitem [{\citenamefont {Reichhardt}\ \emph {et~al.}(2022)\citenamefont {Reichhardt}, \citenamefont {Reichhardt},\ and\ \citenamefont {Milo{\v s}evi{\' c}}}]{reichhardt_statics_2022}%
  \BibitemOpen
  \bibfield  {author} {\bibinfo {author} {\bibfnamefont {C.}~\bibnamefont {Reichhardt}}, \bibinfo {author} {\bibfnamefont {C.~J.~O.}\ \bibnamefont {Reichhardt}}, \ and\ \bibinfo {author} {\bibfnamefont {M.}~\bibnamefont {Milo{\v s}evi{\' c}}},\ }\bibfield  {title} {\enquote {\bibinfo {title} {Statics and dynamics of skyrmions interacting with disorder and nanostructures},}\ }\href {\doibase 10.1103/RevModPhys.94.035005} {\bibfield  {journal} {\bibinfo  {journal} {Rev. Mod. Phys.}\ }\textbf {\bibinfo {volume} {94}},\ \bibinfo {pages} {035005} (\bibinfo {year} {2022})}\BibitemShut {NoStop}%
\bibitem [{\citenamefont {Litzius}\ \emph {et~al.}(2017)\citenamefont {Litzius}, \citenamefont {Lemesh}, \citenamefont {Kr{\" u}ger}, \citenamefont {Bassirian}, \citenamefont {Caretta}, \citenamefont {Richter}, \citenamefont {B{\" u}ttner}, \citenamefont {Sato}, \citenamefont {Tretiakov}, \citenamefont {F{\" o}rster}, \citenamefont {Reeve}, \citenamefont {Weigand}, \citenamefont {Bykova}, \citenamefont {Stoll}, \citenamefont {Sch{\" u}tz}, \citenamefont {Beach},\ and\ \citenamefont {Kl{\" a}ui}}]{litzius_skyrmion_2017}%
  \BibitemOpen
  \bibfield  {author} {\bibinfo {author} {\bibfnamefont {K.}~\bibnamefont {Litzius}}, \bibinfo {author} {\bibfnamefont {I.}~\bibnamefont {Lemesh}}, \bibinfo {author} {\bibfnamefont {B.}~\bibnamefont {Kr{\" u}ger}}, \bibinfo {author} {\bibfnamefont {P.}~\bibnamefont {Bassirian}}, \bibinfo {author} {\bibfnamefont {L.}~\bibnamefont {Caretta}}, \bibinfo {author} {\bibfnamefont {K.}~\bibnamefont {Richter}}, \bibinfo {author} {\bibfnamefont {F.}~\bibnamefont {B{\" u}ttner}}, \bibinfo {author} {\bibfnamefont {K.}~\bibnamefont {Sato}}, \bibinfo {author} {\bibfnamefont {O.~A.}\ \bibnamefont {Tretiakov}}, \bibinfo {author} {\bibfnamefont {J.}~\bibnamefont {F{\" o}rster}}, \bibinfo {author} {\bibfnamefont {R.~M.}\ \bibnamefont {Reeve}}, \bibinfo {author} {\bibfnamefont {M.}~\bibnamefont {Weigand}}, \bibinfo {author} {\bibfnamefont {I.}~\bibnamefont {Bykova}}, \bibinfo {author} {\bibfnamefont {H.}~\bibnamefont {Stoll}}, \bibinfo {author} {\bibfnamefont {G.}~\bibnamefont {Sch{\" u}tz}}, \bibinfo {author} {\bibfnamefont
  {G.~S.~D.}\ \bibnamefont {Beach}}, \ and\ \bibinfo {author} {\bibfnamefont {M.}~\bibnamefont {Kl{\" a}ui}},\ }\bibfield  {title} {\enquote {\bibinfo {title} {Skyrmion {H}all effect revealed by direct time-resolved {X}-ray microscopy},}\ }\href {\doibase 10.1038/NPHYS4000} {\bibfield  {journal} {\bibinfo  {journal} {Nature Phys.}\ }\textbf {\bibinfo {volume} {13}},\ \bibinfo {pages} {170--175} (\bibinfo {year} {2017})}\BibitemShut {NoStop}%
\bibitem [{\citenamefont {Iwasaki}\ \emph {et~al.}(2013{\natexlab{a}})\citenamefont {Iwasaki}, \citenamefont {Mochizuki},\ and\ \citenamefont {Nagaosa}}]{iwasaki_universal_2013}%
  \BibitemOpen
  \bibfield  {author} {\bibinfo {author} {\bibfnamefont {J.}~\bibnamefont {Iwasaki}}, \bibinfo {author} {\bibfnamefont {M.}~\bibnamefont {Mochizuki}}, \ and\ \bibinfo {author} {\bibfnamefont {N.}~\bibnamefont {Nagaosa}},\ }\bibfield  {title} {\enquote {\bibinfo {title} {Universal current-velocity relation of skyrmion motion in chiral magnets},}\ }\href {\doibase 10.1038/ncomms2442} {\bibfield  {journal} {\bibinfo  {journal} {Nature Commun.}\ }\textbf {\bibinfo {volume} {4}},\ \bibinfo {pages} {1463} (\bibinfo {year} {2013}{\natexlab{a}})}\BibitemShut {NoStop}%
\bibitem [{\citenamefont {Jiang}\ \emph {et~al.}(2017)\citenamefont {Jiang}, \citenamefont {Zhang}, \citenamefont {Yu}, \citenamefont {Zhang}, \citenamefont {Wang}, \citenamefont {Jungfleisch}, \citenamefont {Pearson}, \citenamefont {Cheng}, \citenamefont {Heinonen}, \citenamefont {Wang}, \citenamefont {Zhou}, \citenamefont {Hoffmann},\ and\ \citenamefont {te~Velthuis}}]{jiang_direct_2017}%
  \BibitemOpen
  \bibfield  {author} {\bibinfo {author} {\bibfnamefont {W.}~\bibnamefont {Jiang}}, \bibinfo {author} {\bibfnamefont {X.}~\bibnamefont {Zhang}}, \bibinfo {author} {\bibfnamefont {G.}~\bibnamefont {Yu}}, \bibinfo {author} {\bibfnamefont {W.}~\bibnamefont {Zhang}}, \bibinfo {author} {\bibfnamefont {X.}~\bibnamefont {Wang}}, \bibinfo {author} {\bibfnamefont {M.~B.}\ \bibnamefont {Jungfleisch}}, \bibinfo {author} {\bibfnamefont {J.~E.}\ \bibnamefont {Pearson}}, \bibinfo {author} {\bibfnamefont {X.}~\bibnamefont {Cheng}}, \bibinfo {author} {\bibfnamefont {O.}~\bibnamefont {Heinonen}}, \bibinfo {author} {\bibfnamefont {K.~L.}\ \bibnamefont {Wang}}, \bibinfo {author} {\bibfnamefont {Y.}~\bibnamefont {Zhou}}, \bibinfo {author} {\bibfnamefont {A.}~\bibnamefont {Hoffmann}}, \ and\ \bibinfo {author} {\bibfnamefont {S.~G.~E.}\ \bibnamefont {te~Velthuis}},\ }\bibfield  {title} {\enquote {\bibinfo {title} {Direct observation of the skyrmion {H}all effect},}\ }\href {\doibase 10.1038/NPHYS3883} {\bibfield  {journal}
  {\bibinfo  {journal} {Nature Phys.}\ }\textbf {\bibinfo {volume} {13}},\ \bibinfo {pages} {162--169} (\bibinfo {year} {2017})}\BibitemShut {NoStop}%
\bibitem [{\citenamefont {Lin}\ \emph {et~al.}(2013{\natexlab{a}})\citenamefont {Lin}, \citenamefont {Reichhardt}, \citenamefont {Batista},\ and\ \citenamefont {Saxena}}]{lin_driven_2013}%
  \BibitemOpen
  \bibfield  {author} {\bibinfo {author} {\bibfnamefont {S.-Z.}\ \bibnamefont {Lin}}, \bibinfo {author} {\bibfnamefont {C.}~\bibnamefont {Reichhardt}}, \bibinfo {author} {\bibfnamefont {C.~D.}\ \bibnamefont {Batista}}, \ and\ \bibinfo {author} {\bibfnamefont {A.}~\bibnamefont {Saxena}},\ }\bibfield  {title} {\enquote {\bibinfo {title} {Driven skyrmions and dynamical transitions in chiral magnets},}\ }\href {\doibase 10.1103/PhysRevLett.110.207202} {\bibfield  {journal} {\bibinfo  {journal} {Phys. Rev. Lett.}\ }\textbf {\bibinfo {volume} {110}},\ \bibinfo {pages} {207202} (\bibinfo {year} {2013}{\natexlab{a}})}\BibitemShut {NoStop}%
\bibitem [{\citenamefont {Lin}\ \emph {et~al.}(2013{\natexlab{b}})\citenamefont {Lin}, \citenamefont {Reichhardt}, \citenamefont {Batista},\ and\ \citenamefont {Saxena}}]{lin_particle_2013}%
  \BibitemOpen
  \bibfield  {author} {\bibinfo {author} {\bibfnamefont {S.-Z.}\ \bibnamefont {Lin}}, \bibinfo {author} {\bibfnamefont {C.}~\bibnamefont {Reichhardt}}, \bibinfo {author} {\bibfnamefont {C.~D.}\ \bibnamefont {Batista}}, \ and\ \bibinfo {author} {\bibfnamefont {A.}~\bibnamefont {Saxena}},\ }\bibfield  {title} {\enquote {\bibinfo {title} {Particle model for skyrmions in metallic chiral magnets: Dynamics, pinning, and creep},}\ }\href {\doibase 10.1103/PhysRevB.87.214419} {\bibfield  {journal} {\bibinfo  {journal} {Phys. Rev. B}\ }\textbf {\bibinfo {volume} {87}},\ \bibinfo {pages} {214419} (\bibinfo {year} {2013}{\natexlab{b}})}\BibitemShut {NoStop}%
\bibitem [{\citenamefont {Zeissler}\ \emph {et~al.}(2020)\citenamefont {Zeissler}, \citenamefont {Finizio}, \citenamefont {Barton}, \citenamefont {Huxtable}, \citenamefont {Massey}, \citenamefont {Raabe}, \citenamefont {Sadovnikov}, \citenamefont {Nikitov}, \citenamefont {Brearton}, \citenamefont {Hesjedal}, \citenamefont {van~der Laan}, \citenamefont {Rosamond}, \citenamefont {Linfield}, \citenamefont {Burnell},\ and\ \citenamefont {Marrows}}]{zeissler_diameter-independent_2020}%
  \BibitemOpen
  \bibfield  {author} {\bibinfo {author} {\bibfnamefont {Katharina}\ \bibnamefont {Zeissler}}, \bibinfo {author} {\bibfnamefont {Simone}\ \bibnamefont {Finizio}}, \bibinfo {author} {\bibfnamefont {Craig}\ \bibnamefont {Barton}}, \bibinfo {author} {\bibfnamefont {Alexandra~J}\ \bibnamefont {Huxtable}}, \bibinfo {author} {\bibfnamefont {Jamie}\ \bibnamefont {Massey}}, \bibinfo {author} {\bibfnamefont {Jörg}\ \bibnamefont {Raabe}}, \bibinfo {author} {\bibfnamefont {Alexandr~V}\ \bibnamefont {Sadovnikov}}, \bibinfo {author} {\bibfnamefont {Sergey~A}\ \bibnamefont {Nikitov}}, \bibinfo {author} {\bibfnamefont {Richard}\ \bibnamefont {Brearton}}, \bibinfo {author} {\bibfnamefont {Thorsten}\ \bibnamefont {Hesjedal}}, \bibinfo {author} {\bibfnamefont {Gerrit}\ \bibnamefont {van~der Laan}}, \bibinfo {author} {\bibfnamefont {Mark~C}\ \bibnamefont {Rosamond}}, \bibinfo {author} {\bibfnamefont {Edmund~H}\ \bibnamefont {Linfield}}, \bibinfo {author} {\bibfnamefont {Gavin}\ \bibnamefont {Burnell}}, \ and\ \bibinfo {author}
  {\bibfnamefont {Christopher~H}\ \bibnamefont {Marrows}},\ }\bibfield  {title} {\enquote {\bibinfo {title} {Diameter-independent skyrmion {Hall} angle observed in chiral magnetic multilayers},}\ }\href {\doibase 10.1038/s41467-019-14232-9} {\bibfield  {journal} {\bibinfo  {journal} {Nature Communications}\ }\textbf {\bibinfo {volume} {11}},\ \bibinfo {pages} {428} (\bibinfo {year} {2020})}\BibitemShut {NoStop}%
\bibitem [{\citenamefont {Fert}\ \emph {et~al.}(2017)\citenamefont {Fert}, \citenamefont {Reyren},\ and\ \citenamefont {Cros}}]{fert_magnetic_2017}%
  \BibitemOpen
  \bibfield  {author} {\bibinfo {author} {\bibfnamefont {Albert}\ \bibnamefont {Fert}}, \bibinfo {author} {\bibfnamefont {Nicolas}\ \bibnamefont {Reyren}}, \ and\ \bibinfo {author} {\bibfnamefont {Vincent}\ \bibnamefont {Cros}},\ }\bibfield  {title} {\enquote {\bibinfo {title} {Magnetic skyrmions: advances in physics and potential applications},}\ }\href {\doibase 10.1038/natrevmats.2017.31} {\bibfield  {journal} {\bibinfo  {journal} {Nature Reviews Materials}\ }\textbf {\bibinfo {volume} {2}},\ \bibinfo {pages} {1--15} (\bibinfo {year} {2017})}\BibitemShut {NoStop}%
\bibitem [{\citenamefont {Reichhardt}\ \emph {et~al.}(2015{\natexlab{a}})\citenamefont {Reichhardt}, \citenamefont {Ray},\ and\ \citenamefont {Reichhardt}}]{reichhardt_quantized_2015}%
  \BibitemOpen
  \bibfield  {author} {\bibinfo {author} {\bibfnamefont {C.}~\bibnamefont {Reichhardt}}, \bibinfo {author} {\bibfnamefont {D.}~\bibnamefont {Ray}}, \ and\ \bibinfo {author} {\bibfnamefont {C.~J.~Olson}\ \bibnamefont {Reichhardt}},\ }\bibfield  {title} {\enquote {\bibinfo {title} {Quantized transport for a skyrmion moving on a two-dimensional periodic substrate},}\ }\href {\doibase 10.1103/PhysRevB.91.104426} {\bibfield  {journal} {\bibinfo  {journal} {Phys. Rev. B}\ }\textbf {\bibinfo {volume} {91}},\ \bibinfo {pages} {104426} (\bibinfo {year} {2015}{\natexlab{a}})}\BibitemShut {NoStop}%
\bibitem [{\citenamefont {Reichhardt}\ \emph {et~al.}(2018)\citenamefont {Reichhardt}, \citenamefont {Ray},\ and\ \citenamefont {Reichhardt}}]{reichhardt_nonequilibrium_2018}%
  \BibitemOpen
  \bibfield  {author} {\bibinfo {author} {\bibfnamefont {C.}~\bibnamefont {Reichhardt}}, \bibinfo {author} {\bibfnamefont {D.}~\bibnamefont {Ray}}, \ and\ \bibinfo {author} {\bibfnamefont {C.~J.~O.}\ \bibnamefont {Reichhardt}},\ }\bibfield  {title} {\enquote {\bibinfo {title} {Nonequilibrium phases and segregation for skyrmions on periodic pinning arrays},}\ }\href {\doibase 10.1103/PhysRevB.98.134418} {\bibfield  {journal} {\bibinfo  {journal} {Phys. Rev. B}\ }\textbf {\bibinfo {volume} {98}},\ \bibinfo {pages} {134418} (\bibinfo {year} {2018})}\BibitemShut {NoStop}%
\bibitem [{\citenamefont {Feilhauer}\ \emph {et~al.}(2020)\citenamefont {Feilhauer}, \citenamefont {Saha}, \citenamefont {Tobik}, \citenamefont {Zelent}, \citenamefont {Heyderman},\ and\ \citenamefont {Mruczkiewicz}}]{feilhauer_controlled_2020}%
  \BibitemOpen
  \bibfield  {author} {\bibinfo {author} {\bibfnamefont {J.}~\bibnamefont {Feilhauer}}, \bibinfo {author} {\bibfnamefont {S.}~\bibnamefont {Saha}}, \bibinfo {author} {\bibfnamefont {J.}~\bibnamefont {Tobik}}, \bibinfo {author} {\bibfnamefont {M.}~\bibnamefont {Zelent}}, \bibinfo {author} {\bibfnamefont {L.~J.}\ \bibnamefont {Heyderman}}, \ and\ \bibinfo {author} {\bibfnamefont {M.}~\bibnamefont {Mruczkiewicz}},\ }\bibfield  {title} {\enquote {\bibinfo {title} {Controlled motion of skyrmions in a magnetic antidot lattice},}\ }\href {\doibase 10.1103/PhysRevB.102.184425} {\bibfield  {journal} {\bibinfo  {journal} {Phys. Rev. B}\ }\textbf {\bibinfo {volume} {102}},\ \bibinfo {pages} {184425} (\bibinfo {year} {2020})}\BibitemShut {NoStop}%
\bibitem [{\citenamefont {Vizarim}\ \emph {et~al.}(2021{\natexlab{a}})\citenamefont {Vizarim}, \citenamefont {Souza}, \citenamefont {Reichhardt}, \citenamefont {Reichhardt},\ and\ \citenamefont {Venegas}}]{vizarim_directional_2021}%
  \BibitemOpen
  \bibfield  {author} {\bibinfo {author} {\bibfnamefont {N.~P.}\ \bibnamefont {Vizarim}}, \bibinfo {author} {\bibfnamefont {J.~C.~Bellizotti}\ \bibnamefont {Souza}}, \bibinfo {author} {\bibfnamefont {C.}~\bibnamefont {Reichhardt}}, \bibinfo {author} {\bibfnamefont {C.~J.~O.}\ \bibnamefont {Reichhardt}}, \ and\ \bibinfo {author} {\bibfnamefont {P.~A.}\ \bibnamefont {Venegas}},\ }\bibfield  {title} {\enquote {\bibinfo {title} {Directional locking and the influence of obstacle density on skyrmion dynamics in triangular and honeycomb arrays},}\ }\href {\doibase 10.1088/1361-648X/ac0081} {\bibfield  {journal} {\bibinfo  {journal} {J. Phys: Condens. Matter}\ }\textbf {\bibinfo {volume} {33}},\ \bibinfo {pages} {305801} (\bibinfo {year} {2021}{\natexlab{a}})}\BibitemShut {NoStop}%
\bibitem [{\citenamefont {Vizarim}\ \emph {et~al.}(2020{\natexlab{a}})\citenamefont {Vizarim}, \citenamefont {Reichhardt}, \citenamefont {Reichhardt},\ and\ \citenamefont {Venegas}}]{vizarim_skyrmion_2020}%
  \BibitemOpen
  \bibfield  {author} {\bibinfo {author} {\bibfnamefont {N.~P.}\ \bibnamefont {Vizarim}}, \bibinfo {author} {\bibfnamefont {C.}~\bibnamefont {Reichhardt}}, \bibinfo {author} {\bibfnamefont {C.~J.~O.}\ \bibnamefont {Reichhardt}}, \ and\ \bibinfo {author} {\bibfnamefont {P.~A.}\ \bibnamefont {Venegas}},\ }\bibfield  {title} {\enquote {\bibinfo {title} {Skyrmion dynamics and topological sorting on periodic obstacle arrays},}\ }\href {\doibase 10.1088/1367-2630/ab8045} {\bibfield  {journal} {\bibinfo  {journal} {New J. Phys.}\ }\textbf {\bibinfo {volume} {22}},\ \bibinfo {pages} {53025} (\bibinfo {year} {2020}{\natexlab{a}})}\BibitemShut {NoStop}%
\bibitem [{\citenamefont {Reichhardt}\ and\ \citenamefont {Reichhardt}(2022)}]{reichhardt_commensuration_2022}%
  \BibitemOpen
  \bibfield  {author} {\bibinfo {author} {\bibfnamefont {C.}~\bibnamefont {Reichhardt}}\ and\ \bibinfo {author} {\bibfnamefont {C.~J.~O.}\ \bibnamefont {Reichhardt}},\ }\bibfield  {title} {\enquote {\bibinfo {title} {Commensuration effects on skyrmion {Hall} angle and drag for manipulation of skyrmions on two-dimensional periodic substrates},}\ }\href {\doibase 10.1103/PhysRevB.105.214437} {\bibfield  {journal} {\bibinfo  {journal} {Phys. Rev. B}\ }\textbf {\bibinfo {volume} {105}},\ \bibinfo {pages} {214437} (\bibinfo {year} {2022})}\BibitemShut {NoStop}%
\bibitem [{\citenamefont {Carvalho-Santos}\ \emph {et~al.}(2021)\citenamefont {Carvalho-Santos}, \citenamefont {Castro}, \citenamefont {Salazar-Aravena}, \citenamefont {Laroze}, \citenamefont {Corona}, \citenamefont {Allende},\ and\ \citenamefont {Altbir}}]{carvalho-santos_skyrmion_2021}%
  \BibitemOpen
  \bibfield  {author} {\bibinfo {author} {\bibfnamefont {V.~L.}\ \bibnamefont {Carvalho-Santos}}, \bibinfo {author} {\bibfnamefont {M.~A.}\ \bibnamefont {Castro}}, \bibinfo {author} {\bibfnamefont {D.}~\bibnamefont {Salazar-Aravena}}, \bibinfo {author} {\bibfnamefont {D.}~\bibnamefont {Laroze}}, \bibinfo {author} {\bibfnamefont {R.~M.}\ \bibnamefont {Corona}}, \bibinfo {author} {\bibfnamefont {S.}~\bibnamefont {Allende}}, \ and\ \bibinfo {author} {\bibfnamefont {D.}~\bibnamefont {Altbir}},\ }\bibfield  {title} {\enquote {\bibinfo {title} {Skyrmion propagation along curved racetracks},}\ }\href {\doibase 10.1063/5.0045969} {\bibfield  {journal} {\bibinfo  {journal} {Appl. Phys. Lett.}\ }\textbf {\bibinfo {volume} {118}},\ \bibinfo {pages} {172407} (\bibinfo {year} {2021})}\BibitemShut {NoStop}%
\bibitem [{\citenamefont {Korniienko}\ \emph {et~al.}(2020)\citenamefont {Korniienko}, \citenamefont {K{\' a}kay}, \citenamefont {Sheka},\ and\ \citenamefont {Kravchuk}}]{korniienko_effect_2020}%
  \BibitemOpen
  \bibfield  {author} {\bibinfo {author} {\bibfnamefont {A.}~\bibnamefont {Korniienko}}, \bibinfo {author} {\bibfnamefont {A.}~\bibnamefont {K{\' a}kay}}, \bibinfo {author} {\bibfnamefont {D.~D.}\ \bibnamefont {Sheka}}, \ and\ \bibinfo {author} {\bibfnamefont {V.~P.}\ \bibnamefont {Kravchuk}},\ }\bibfield  {title} {\enquote {\bibinfo {title} {Effect of curvature on the eigenstates of magnetic skyrmions},}\ }\href {\doibase 10.1103/PhysRevB.102.014432} {\bibfield  {journal} {\bibinfo  {journal} {Phys. Rev. B}\ }\textbf {\bibinfo {volume} {102}},\ \bibinfo {pages} {014432} (\bibinfo {year} {2020})}\BibitemShut {NoStop}%
\bibitem [{\citenamefont {Yershov}\ \emph {et~al.}(2022)\citenamefont {Yershov}, \citenamefont {K{\' a}kay},\ and\ \citenamefont {Kravchuk}}]{yershov_curvature-induced_2022}%
  \BibitemOpen
  \bibfield  {author} {\bibinfo {author} {\bibfnamefont {K.~V.}\ \bibnamefont {Yershov}}, \bibinfo {author} {\bibfnamefont {A.}~\bibnamefont {K{\' a}kay}}, \ and\ \bibinfo {author} {\bibfnamefont {V.~P.}\ \bibnamefont {Kravchuk}},\ }\bibfield  {title} {\enquote {\bibinfo {title} {Curvature-induced drift and deformation of magnetic skyrmions: Comparison of the ferromagnetic and antiferromagnetic cases},}\ }\href {\doibase 10.1103/PhysRevB.105.054425} {\bibfield  {journal} {\bibinfo  {journal} {Phys. Rev. B}\ }\textbf {\bibinfo {volume} {105}},\ \bibinfo {pages} {054425} (\bibinfo {year} {2022})}\BibitemShut {NoStop}%
\bibitem [{\citenamefont {Vizarim}\ \emph {et~al.}(2021{\natexlab{b}})\citenamefont {Vizarim}, \citenamefont {Reichhardt}, \citenamefont {Venegas},\ and\ \citenamefont {Reichhardt}}]{vizarim_guided_2021}%
  \BibitemOpen
  \bibfield  {author} {\bibinfo {author} {\bibfnamefont {N.~P.}\ \bibnamefont {Vizarim}}, \bibinfo {author} {\bibfnamefont {C.}~\bibnamefont {Reichhardt}}, \bibinfo {author} {\bibfnamefont {P.~A.}\ \bibnamefont {Venegas}}, \ and\ \bibinfo {author} {\bibfnamefont {C.~J.~O.}\ \bibnamefont {Reichhardt}},\ }\bibfield  {title} {\enquote {\bibinfo {title} {Guided skyrmion motion along pinning array interfaces},}\ }\href {\doibase 10.1016/j.jmmm.2020.167710} {\bibfield  {journal} {\bibinfo  {journal} {J. Mag. Mag. Mater.}\ }\textbf {\bibinfo {volume} {528}},\ \bibinfo {pages} {167710} (\bibinfo {year} {2021}{\natexlab{b}})}\BibitemShut {NoStop}%
\bibitem [{\citenamefont {Zhang}\ \emph {et~al.}(2022{\natexlab{a}})\citenamefont {Zhang}, \citenamefont {Wang}, \citenamefont {Song}, \citenamefont {Mehmood}, \citenamefont {Zeng}, \citenamefont {Ma}, \citenamefont {Wang},\ and\ \citenamefont {Liu}}]{zhang_edge-guided_2022}%
  \BibitemOpen
  \bibfield  {author} {\bibinfo {author} {\bibfnamefont {C.-L.}\ \bibnamefont {Zhang}}, \bibinfo {author} {\bibfnamefont {J.-N.}\ \bibnamefont {Wang}}, \bibinfo {author} {\bibfnamefont {C.-K.}\ \bibnamefont {Song}}, \bibinfo {author} {\bibfnamefont {N.}~\bibnamefont {Mehmood}}, \bibinfo {author} {\bibfnamefont {Z.-Z.}\ \bibnamefont {Zeng}}, \bibinfo {author} {\bibfnamefont {Y.-X.}\ \bibnamefont {Ma}}, \bibinfo {author} {\bibfnamefont {J.-B.}\ \bibnamefont {Wang}}, \ and\ \bibinfo {author} {\bibfnamefont {Q.-F.}\ \bibnamefont {Liu}},\ }\bibfield  {title} {\enquote {\bibinfo {title} {Edge-guided heart-shaped skyrmion},}\ }\href {\doibase 10.1007/s12598-021-01844-8} {\bibfield  {journal} {\bibinfo  {journal} {Rare Metals}\ }\textbf {\bibinfo {volume} {41}},\ \bibinfo {pages} {865--870} (\bibinfo {year} {2022}{\natexlab{a}})}\BibitemShut {NoStop}%
\bibitem [{\citenamefont {Reichhardt}\ \emph {et~al.}(2015{\natexlab{b}})\citenamefont {Reichhardt}, \citenamefont {Ray},\ and\ \citenamefont {Reichhardt}}]{reichhardt_magnus-induced_2015}%
  \BibitemOpen
  \bibfield  {author} {\bibinfo {author} {\bibfnamefont {C.}~\bibnamefont {Reichhardt}}, \bibinfo {author} {\bibfnamefont {D.}~\bibnamefont {Ray}}, \ and\ \bibinfo {author} {\bibfnamefont {C.~J.~Olson}\ \bibnamefont {Reichhardt}},\ }\bibfield  {title} {\enquote {\bibinfo {title} {Magnus-induced ratchet effects for skyrmions interacting with asymmetric substrates},}\ }\href {\doibase 10.1088/1367-2630/17/7/073034} {\bibfield  {journal} {\bibinfo  {journal} {New J. Phys.}\ }\textbf {\bibinfo {volume} {17}},\ \bibinfo {pages} {73034} (\bibinfo {year} {2015}{\natexlab{b}})}\BibitemShut {NoStop}%
\bibitem [{\citenamefont {Souza}\ \emph {et~al.}(2021)\citenamefont {Souza}, \citenamefont {Vizarim}, \citenamefont {Reichhardt}, \citenamefont {Reichhardt},\ and\ \citenamefont {Venegas}}]{souza_skyrmion_2021}%
  \BibitemOpen
  \bibfield  {author} {\bibinfo {author} {\bibfnamefont {J.~C.~Bellizotti}\ \bibnamefont {Souza}}, \bibinfo {author} {\bibfnamefont {N.~P.}\ \bibnamefont {Vizarim}}, \bibinfo {author} {\bibfnamefont {C.~J.~O.}\ \bibnamefont {Reichhardt}}, \bibinfo {author} {\bibfnamefont {C.}~\bibnamefont {Reichhardt}}, \ and\ \bibinfo {author} {\bibfnamefont {P.~A.}\ \bibnamefont {Venegas}},\ }\bibfield  {title} {\enquote {\bibinfo {title} {Skyrmion ratchet in funnel geometries},}\ }\href {\doibase 10.1103/PhysRevB.104.054434} {\bibfield  {journal} {\bibinfo  {journal} {Phys. Rev. B}\ }\textbf {\bibinfo {volume} {104}},\ \bibinfo {pages} {54434} (\bibinfo {year} {2021})}\BibitemShut {NoStop}%
\bibitem [{\citenamefont {Chen}\ \emph {et~al.}(2019)\citenamefont {Chen}, \citenamefont {Liu}, \citenamefont {Ji},\ and\ \citenamefont {Zheng}}]{chen_skyrmion_2019}%
  \BibitemOpen
  \bibfield  {author} {\bibinfo {author} {\bibfnamefont {W.}~\bibnamefont {Chen}}, \bibinfo {author} {\bibfnamefont {L.}~\bibnamefont {Liu}}, \bibinfo {author} {\bibfnamefont {Y.}~\bibnamefont {Ji}}, \ and\ \bibinfo {author} {\bibfnamefont {Y.}~\bibnamefont {Zheng}},\ }\bibfield  {title} {\enquote {\bibinfo {title} {Skyrmion ratchet effect driven by a biharmonic force},}\ }\href {\doibase 10.1103/PhysRevB.99.064431} {\bibfield  {journal} {\bibinfo  {journal} {Phys. Rev. B}\ }\textbf {\bibinfo {volume} {99}},\ \bibinfo {pages} {64431} (\bibinfo {year} {2019})}\BibitemShut {NoStop}%
\bibitem [{\citenamefont {G{\" o}bel}\ and\ \citenamefont {Mertig}(2021)}]{gobel_skyrmion_2021}%
  \BibitemOpen
  \bibfield  {author} {\bibinfo {author} {\bibfnamefont {B.}~\bibnamefont {G{\" o}bel}}\ and\ \bibinfo {author} {\bibfnamefont {I.}~\bibnamefont {Mertig}},\ }\bibfield  {title} {\enquote {\bibinfo {title} {Skyrmion ratchet propagation: utilizing the skyrmion {Hall} effect in {AC} racetrack storage devices},}\ }\href {\doibase 10.1038/s41598-021-81992-0} {\bibfield  {journal} {\bibinfo  {journal} {Sci. Rep.}\ }\textbf {\bibinfo {volume} {11}},\ \bibinfo {pages} {3020} (\bibinfo {year} {2021})}\BibitemShut {NoStop}%
\bibitem [{\citenamefont {Souza}\ \emph {et~al.}(2024)\citenamefont {Souza}, \citenamefont {Vizarim}, \citenamefont {Reichhardt}, \citenamefont {Reichhardt},\ and\ \citenamefont {Venegas}}]{souza_controlled_2024}%
  \BibitemOpen
  \bibfield  {author} {\bibinfo {author} {\bibfnamefont {J.~C.~Bellizotti}\ \bibnamefont {Souza}}, \bibinfo {author} {\bibfnamefont {N.~P.}\ \bibnamefont {Vizarim}}, \bibinfo {author} {\bibfnamefont {C.~J.~O.}\ \bibnamefont {Reichhardt}}, \bibinfo {author} {\bibfnamefont {C.}~\bibnamefont {Reichhardt}}, \ and\ \bibinfo {author} {\bibfnamefont {P.~A.}\ \bibnamefont {Venegas}},\ }\bibfield  {title} {\enquote {\bibinfo {title} {Controlled skyrmion ratchet in linear protrusion defects},}\ }\href {\doibase 10.1103/PhysRevB.109.054407} {\bibfield  {journal} {\bibinfo  {journal} {Phys. Rev. B}\ }\textbf {\bibinfo {volume} {109}},\ \bibinfo {pages} {054407} (\bibinfo {year} {2024})}\BibitemShut {NoStop}%
\bibitem [{\citenamefont {Yanes}\ \emph {et~al.}(2019)\citenamefont {Yanes}, \citenamefont {Garcia-Sanchez}, \citenamefont {Luis}, \citenamefont {Martinez}, \citenamefont {Raposo}, \citenamefont {Torres},\ and\ \citenamefont {Lopez-Diaz}}]{yanes_skyrmion_2019}%
  \BibitemOpen
  \bibfield  {author} {\bibinfo {author} {\bibfnamefont {R.}~\bibnamefont {Yanes}}, \bibinfo {author} {\bibfnamefont {F.}~\bibnamefont {Garcia-Sanchez}}, \bibinfo {author} {\bibfnamefont {R.~F.}\ \bibnamefont {Luis}}, \bibinfo {author} {\bibfnamefont {E.}~\bibnamefont {Martinez}}, \bibinfo {author} {\bibfnamefont {V.}~\bibnamefont {Raposo}}, \bibinfo {author} {\bibfnamefont {L.}~\bibnamefont {Torres}}, \ and\ \bibinfo {author} {\bibfnamefont {L.}~\bibnamefont {Lopez-Diaz}},\ }\bibfield  {title} {\enquote {\bibinfo {title} {Skyrmion motion induced by voltage-controlled in-plane strain gradients},}\ }\href {\doibase 10.1063/1.5119085} {\bibfield  {journal} {\bibinfo  {journal} {Appl. Phys. Lett.}\ }\textbf {\bibinfo {volume} {115}},\ \bibinfo {pages} {132401} (\bibinfo {year} {2019})}\BibitemShut {NoStop}%
\bibitem [{\citenamefont {Zhang}\ \emph {et~al.}(2018)\citenamefont {Zhang}, \citenamefont {Wang}, \citenamefont {Burn}, \citenamefont {Peng}, \citenamefont {Berger}, \citenamefont {Bauer}, \citenamefont {Pfleiderer}, \citenamefont {van~der Laan},\ and\ \citenamefont {Hesjedal}}]{zhang_manipulation_2018}%
  \BibitemOpen
  \bibfield  {author} {\bibinfo {author} {\bibfnamefont {S.~L.}\ \bibnamefont {Zhang}}, \bibinfo {author} {\bibfnamefont {W.~W.}\ \bibnamefont {Wang}}, \bibinfo {author} {\bibfnamefont {D.~M.}\ \bibnamefont {Burn}}, \bibinfo {author} {\bibfnamefont {H.}~\bibnamefont {Peng}}, \bibinfo {author} {\bibfnamefont {H.}~\bibnamefont {Berger}}, \bibinfo {author} {\bibfnamefont {A.}~\bibnamefont {Bauer}}, \bibinfo {author} {\bibfnamefont {C.}~\bibnamefont {Pfleiderer}}, \bibinfo {author} {\bibfnamefont {G.}~\bibnamefont {van~der Laan}}, \ and\ \bibinfo {author} {\bibfnamefont {T.}~\bibnamefont {Hesjedal}},\ }\bibfield  {title} {\enquote {\bibinfo {title} {Manipulation of skyrmion motion by magnetic field gradients},}\ }\href {\doibase 10.1038/s41467-018-04563-4} {\bibfield  {journal} {\bibinfo  {journal} {Nature Commun.}\ }\textbf {\bibinfo {volume} {9}},\ \bibinfo {pages} {2115} (\bibinfo {year} {2018})}\BibitemShut {NoStop}%
\bibitem [{\citenamefont {Everschor}\ \emph {et~al.}(2012)\citenamefont {Everschor}, \citenamefont {Garst}, \citenamefont {Binz}, \citenamefont {Jonietz}, \citenamefont {M{\" u}hlbauer}, \citenamefont {Pfleiderer},\ and\ \citenamefont {Rosch}}]{everschor_rotating_2012}%
  \BibitemOpen
  \bibfield  {author} {\bibinfo {author} {\bibfnamefont {K.}~\bibnamefont {Everschor}}, \bibinfo {author} {\bibfnamefont {M.}~\bibnamefont {Garst}}, \bibinfo {author} {\bibfnamefont {B.}~\bibnamefont {Binz}}, \bibinfo {author} {\bibfnamefont {F.}~\bibnamefont {Jonietz}}, \bibinfo {author} {\bibfnamefont {S.}~\bibnamefont {M{\" u}hlbauer}}, \bibinfo {author} {\bibfnamefont {C.}~\bibnamefont {Pfleiderer}}, \ and\ \bibinfo {author} {\bibfnamefont {A.}~\bibnamefont {Rosch}},\ }\bibfield  {title} {\enquote {\bibinfo {title} {Rotating skyrmion lattices by spin torques and field or temperature gradients},}\ }\href {\doibase 10.1103/PhysRevB.86.054432} {\bibfield  {journal} {\bibinfo  {journal} {Phys. Rev. B}\ }\textbf {\bibinfo {volume} {86}},\ \bibinfo {pages} {54432} (\bibinfo {year} {2012})}\BibitemShut {NoStop}%
\bibitem [{\citenamefont {Kong}\ and\ \citenamefont {Zang}(2013)}]{kong_dynamics_2013}%
  \BibitemOpen
  \bibfield  {author} {\bibinfo {author} {\bibfnamefont {L.}~\bibnamefont {Kong}}\ and\ \bibinfo {author} {\bibfnamefont {J.}~\bibnamefont {Zang}},\ }\bibfield  {title} {\enquote {\bibinfo {title} {Dynamics of an insulating skyrmion under a temperature gradient},}\ }\href {\doibase 10.1103/PhysRevLett.111.067203} {\bibfield  {journal} {\bibinfo  {journal} {Phys. Rev. Lett.}\ }\textbf {\bibinfo {volume} {111}},\ \bibinfo {pages} {67203} (\bibinfo {year} {2013})}\BibitemShut {NoStop}%
\bibitem [{\citenamefont {Gong}\ \emph {et~al.}(2020)\citenamefont {Gong}, \citenamefont {Yuan},\ and\ \citenamefont {Wang}}]{gong_current-driven_2020}%
  \BibitemOpen
  \bibfield  {author} {\bibinfo {author} {\bibfnamefont {X.}~\bibnamefont {Gong}}, \bibinfo {author} {\bibfnamefont {H.~Y.}\ \bibnamefont {Yuan}}, \ and\ \bibinfo {author} {\bibfnamefont {X.~R.}\ \bibnamefont {Wang}},\ }\bibfield  {title} {\enquote {\bibinfo {title} {Current-driven skyrmion motion in granular films},}\ }\href {\doibase 10.1103/PhysRevB.101.064421} {\bibfield  {journal} {\bibinfo  {journal} {Phys. Rev. B}\ }\textbf {\bibinfo {volume} {101}},\ \bibinfo {pages} {064421} (\bibinfo {year} {2020})}\BibitemShut {NoStop}%
\bibitem [{\citenamefont {Del-Valle}\ \emph {et~al.}(2022)\citenamefont {Del-Valle}, \citenamefont {Castell-Queralt}, \citenamefont {Gonz{\' a}lez-G{\' o}mez},\ and\ \citenamefont {Navau}}]{del-valle_defect_2022}%
  \BibitemOpen
  \bibfield  {author} {\bibinfo {author} {\bibfnamefont {N.}~\bibnamefont {Del-Valle}}, \bibinfo {author} {\bibfnamefont {J.}~\bibnamefont {Castell-Queralt}}, \bibinfo {author} {\bibfnamefont {L.}~\bibnamefont {Gonz{\' a}lez-G{\' o}mez}}, \ and\ \bibinfo {author} {\bibfnamefont {C.}~\bibnamefont {Navau}},\ }\bibfield  {title} {\enquote {\bibinfo {title} {Defect modeling in skyrmionic ferromagnetic systems},}\ }\href {\doibase 10.1063/5.0072709} {\bibfield  {journal} {\bibinfo  {journal} {APL Mater.}\ }\textbf {\bibinfo {volume} {10}},\ \bibinfo {pages} {10702} (\bibinfo {year} {2022})}\BibitemShut {NoStop}%
\bibitem [{\citenamefont {Yuan}\ \emph {et~al.}(2019)\citenamefont {Yuan}, \citenamefont {Wang}, \citenamefont {Yung},\ and\ \citenamefont {Wang}}]{yuan_wiggling_2019}%
  \BibitemOpen
  \bibfield  {author} {\bibinfo {author} {\bibfnamefont {H.~Y.}\ \bibnamefont {Yuan}}, \bibinfo {author} {\bibfnamefont {X.~S.}\ \bibnamefont {Wang}}, \bibinfo {author} {\bibfnamefont {Man-Hong}\ \bibnamefont {Yung}}, \ and\ \bibinfo {author} {\bibfnamefont {X.~R.}\ \bibnamefont {Wang}},\ }\bibfield  {title} {\enquote {\bibinfo {title} {Wiggling skyrmion propagation under parametric pumping},}\ }\href {\doibase 10.1103/PhysRevB.99.014428} {\bibfield  {journal} {\bibinfo  {journal} {Phys. Rev. B}\ }\textbf {\bibinfo {volume} {99}},\ \bibinfo {pages} {014428} (\bibinfo {year} {2019})}\BibitemShut {NoStop}%
\bibitem [{\citenamefont {Zhang}\ \emph {et~al.}(2015)\citenamefont {Zhang}, \citenamefont {Zhou}, \citenamefont {Ezawa}, \citenamefont {Zhao},\ and\ \citenamefont {Zhao}}]{zhang_magnetic_2015}%
  \BibitemOpen
  \bibfield  {author} {\bibinfo {author} {\bibfnamefont {X.}~\bibnamefont {Zhang}}, \bibinfo {author} {\bibfnamefont {Y.}~\bibnamefont {Zhou}}, \bibinfo {author} {\bibfnamefont {M.}~\bibnamefont {Ezawa}}, \bibinfo {author} {\bibfnamefont {G.~P.}\ \bibnamefont {Zhao}}, \ and\ \bibinfo {author} {\bibfnamefont {W.}~\bibnamefont {Zhao}},\ }\bibfield  {title} {\enquote {\bibinfo {title} {Magnetic skyrmion transistor: skyrmion motion in a voltage-gated nanotrack},}\ }\href {\doibase 10.1038/srep11369} {\bibfield  {journal} {\bibinfo  {journal} {Sci. Rep.}\ }\textbf {\bibinfo {volume} {5}},\ \bibinfo {pages} {11369} (\bibinfo {year} {2015})}\BibitemShut {NoStop}%
\bibitem [{\citenamefont {Zhao}\ \emph {et~al.}(2020)\citenamefont {Zhao}, \citenamefont {Liang}, \citenamefont {Xia}, \citenamefont {Zhao},\ and\ \citenamefont {Zhou}}]{zhao_ferromagnetic_2020}%
  \BibitemOpen
  \bibfield  {author} {\bibinfo {author} {\bibfnamefont {L.}~\bibnamefont {Zhao}}, \bibinfo {author} {\bibfnamefont {X.}~\bibnamefont {Liang}}, \bibinfo {author} {\bibfnamefont {J.}~\bibnamefont {Xia}}, \bibinfo {author} {\bibfnamefont {G.}~\bibnamefont {Zhao}}, \ and\ \bibinfo {author} {\bibfnamefont {Y.}~\bibnamefont {Zhou}},\ }\bibfield  {title} {\enquote {\bibinfo {title} {A ferromagnetic skyrmion-based diode with a voltage-controlled potential barrier},}\ }\href {\doibase 10.1039/C9NR10528J} {\bibfield  {journal} {\bibinfo  {journal} {Nanoscale}\ }\textbf {\bibinfo {volume} {17}},\ \bibinfo {pages} {9507} (\bibinfo {year} {2020})}\BibitemShut {NoStop}%
\bibitem [{\citenamefont {Menezes}\ \emph {et~al.}(2019)\citenamefont {Menezes}, \citenamefont {Neto}, \citenamefont {Silva},\ and\ \citenamefont {Milo{\v s}evi{\' c}}}]{menezes_manipulation_2019}%
  \BibitemOpen
  \bibfield  {author} {\bibinfo {author} {\bibfnamefont {R.~M.}\ \bibnamefont {Menezes}}, \bibinfo {author} {\bibfnamefont {J.~F.~S.}\ \bibnamefont {Neto}}, \bibinfo {author} {\bibfnamefont {C.~C. de~Souza}\ \bibnamefont {Silva}}, \ and\ \bibinfo {author} {\bibfnamefont {M.~V.}\ \bibnamefont {Milo{\v s}evi{\' c}}},\ }\bibfield  {title} {\enquote {\bibinfo {title} {Manipulation of magnetic skyrmions by superconducting vortices in ferromagnet-superconductor heterostructures},}\ }\href {\doibase 10.1103/PhysRevB.100.014431} {\bibfield  {journal} {\bibinfo  {journal} {Phys. Rev. B}\ }\textbf {\bibinfo {volume} {100}},\ \bibinfo {pages} {14431} (\bibinfo {year} {2019})}\BibitemShut {NoStop}%
\bibitem [{\citenamefont {Neto}\ and\ \citenamefont {Silva}(2022)}]{neto_mesoscale_2022}%
  \BibitemOpen
  \bibfield  {author} {\bibinfo {author} {\bibfnamefont {J.~F.}\ \bibnamefont {Neto}}\ and\ \bibinfo {author} {\bibfnamefont {C.~C. de~Souza}\ \bibnamefont {Silva}},\ }\bibfield  {title} {\enquote {\bibinfo {title} {Mesoscale phase separation of skyrmion-vortex matter in chiral-magnet--superconductor heterostructures},}\ }\href {\doibase 10.1103/PhysRevLett.128.057001} {\bibfield  {journal} {\bibinfo  {journal} {Phys. Rev. Lett.}\ }\textbf {\bibinfo {volume} {128}},\ \bibinfo {pages} {057001} (\bibinfo {year} {2022})}\BibitemShut {NoStop}%
\bibitem [{\citenamefont {Zhang}\ \emph {et~al.}(2022{\natexlab{b}})\citenamefont {Zhang}, \citenamefont {Xia},\ and\ \citenamefont {Liu}}]{zhang_structural_2022}%
  \BibitemOpen
  \bibfield  {author} {\bibinfo {author} {\bibfnamefont {X.}~\bibnamefont {Zhang}}, \bibinfo {author} {\bibfnamefont {J.}~\bibnamefont {Xia}}, \ and\ \bibinfo {author} {\bibfnamefont {X.}~\bibnamefont {Liu}},\ }\bibfield  {title} {\enquote {\bibinfo {title} {Structural transition of skyrmion quasiparticles under compression},}\ }\href {\doibase 10.1103/PhysRevB.105.184402} {\bibfield  {journal} {\bibinfo  {journal} {Phys. Rev. B}\ }\textbf {\bibinfo {volume} {105}},\ \bibinfo {pages} {184402} (\bibinfo {year} {2022}{\natexlab{b}})}\BibitemShut {NoStop}%
\bibitem [{\citenamefont {Bellizotti~Souza}\ \emph {et~al.}(2023)\citenamefont {Bellizotti~Souza}, \citenamefont {Vizarim}, \citenamefont {Reichhardt}, \citenamefont {Reichhardt},\ and\ \citenamefont {Venegas}}]{bellizotti_souza_spontaneous_2023}%
  \BibitemOpen
  \bibfield  {author} {\bibinfo {author} {\bibfnamefont {J.~C.}\ \bibnamefont {Bellizotti~Souza}}, \bibinfo {author} {\bibfnamefont {N.~P.}\ \bibnamefont {Vizarim}}, \bibinfo {author} {\bibfnamefont {C.~J.~O.}\ \bibnamefont {Reichhardt}}, \bibinfo {author} {\bibfnamefont {C.}~\bibnamefont {Reichhardt}}, \ and\ \bibinfo {author} {\bibfnamefont {P.~A.}\ \bibnamefont {Venegas}},\ }\bibfield  {title} {\enquote {\bibinfo {title} {Spontaneous skyrmion conformal lattice and transverse motion during dc and ac compression},}\ }\href {\doibase 10.1088/1367-2630/acd46f} {\bibfield  {journal} {\bibinfo  {journal} {New J. Phys.}\ }\textbf {\bibinfo {volume} {25}},\ \bibinfo {pages} {053020} (\bibinfo {year} {2023})}\BibitemShut {NoStop}%
\bibitem [{\citenamefont {Zhang}\ \emph {et~al.}(2023)\citenamefont {Zhang}, \citenamefont {Xia}, \citenamefont {Tretiakov}, \citenamefont {Ezawa}, \citenamefont {Zhao}, \citenamefont {Zhou}, \citenamefont {Liu},\ and\ \citenamefont {Mochizuki}}]{zhang_laminar_2023}%
  \BibitemOpen
  \bibfield  {author} {\bibinfo {author} {\bibfnamefont {X.}~\bibnamefont {Zhang}}, \bibinfo {author} {\bibfnamefont {J.}~\bibnamefont {Xia}}, \bibinfo {author} {\bibfnamefont {O.~A.}\ \bibnamefont {Tretiakov}}, \bibinfo {author} {\bibfnamefont {M.}~\bibnamefont {Ezawa}}, \bibinfo {author} {\bibfnamefont {G.}~\bibnamefont {Zhao}}, \bibinfo {author} {\bibfnamefont {Y.}~\bibnamefont {Zhou}}, \bibinfo {author} {\bibfnamefont {X.}~\bibnamefont {Liu}}, \ and\ \bibinfo {author} {\bibfnamefont {M.}~\bibnamefont {Mochizuki}},\ }\bibfield  {title} {\enquote {\bibinfo {title} {Laminar and transiently disordered dynamics of magnetic-skyrmion pipe flow},}\ }\href {\doibase 10.1103/PhysRevB.108.144428} {\bibfield  {journal} {\bibinfo  {journal} {Phys. Rev. B}\ }\textbf {\bibinfo {volume} {108}},\ \bibinfo {pages} {144428} (\bibinfo {year} {2023})}\BibitemShut {NoStop}%
\bibitem [{\citenamefont {Vizarim}\ \emph {et~al.}(2022)\citenamefont {Vizarim}, \citenamefont {Souza}, \citenamefont {Reichhardt}, \citenamefont {Reichhardt}, \citenamefont {Milo{\v s}evi{\' c}},\ and\ \citenamefont {Venegas}}]{vizarim_soliton_2022}%
  \BibitemOpen
  \bibfield  {author} {\bibinfo {author} {\bibfnamefont {N.~P.}\ \bibnamefont {Vizarim}}, \bibinfo {author} {\bibfnamefont {J.~C.~Bellizotti}\ \bibnamefont {Souza}}, \bibinfo {author} {\bibfnamefont {C.~J.~O.}\ \bibnamefont {Reichhardt}}, \bibinfo {author} {\bibfnamefont {C.}~\bibnamefont {Reichhardt}}, \bibinfo {author} {\bibfnamefont {M.~V.}\ \bibnamefont {Milo{\v s}evi{\' c}}}, \ and\ \bibinfo {author} {\bibfnamefont {P.~A.}\ \bibnamefont {Venegas}},\ }\bibfield  {title} {\enquote {\bibinfo {title} {Soliton motion in skyrmion chains: Stabilization and guidance by nanoengineered pinning},}\ }\href {\doibase 10.1103/PhysRevB.105.224409} {\bibfield  {journal} {\bibinfo  {journal} {Phys. Rev. B}\ }\textbf {\bibinfo {volume} {105}},\ \bibinfo {pages} {224409} (\bibinfo {year} {2022})}\BibitemShut {NoStop}%
\bibitem [{\citenamefont {Souza}\ \emph {et~al.}(2023)\citenamefont {Souza}, \citenamefont {Vizarim}, \citenamefont {Reichhardt}, \citenamefont {Reichhardt},\ and\ \citenamefont {Venegas}}]{souza_soliton_2023}%
  \BibitemOpen
  \bibfield  {author} {\bibinfo {author} {\bibfnamefont {J.~C.~Bellizotti}\ \bibnamefont {Souza}}, \bibinfo {author} {\bibfnamefont {N.~P.}\ \bibnamefont {Vizarim}}, \bibinfo {author} {\bibfnamefont {C.~J.~O.}\ \bibnamefont {Reichhardt}}, \bibinfo {author} {\bibfnamefont {C.}~\bibnamefont {Reichhardt}}, \ and\ \bibinfo {author} {\bibfnamefont {P.~A.}\ \bibnamefont {Venegas}},\ }\bibfield  {title} {\enquote {\bibinfo {title} {Soliton motion induced along ferromagnetic skyrmion chains in chiral thin nanotracks},}\ }\href {\doibase 10.1016/j.jmmm.2023.171280} {\bibfield  {journal} {\bibinfo  {journal} {J. Mag. Mag. Mater.}\ }\textbf {\bibinfo {volume} {587}},\ \bibinfo {pages} {171280} (\bibinfo {year} {2023})}\BibitemShut {NoStop}%
\bibitem [{\citenamefont {Du}\ \emph {et~al.}(2024)\citenamefont {Du}, \citenamefont {Song}, \citenamefont {Wang}, \citenamefont {Zhang}, \citenamefont {Wang}, \citenamefont {Zheng}, \citenamefont {Tian}, \citenamefont {Dunin-Borkowski},\ and\ \citenamefont {Zang}}]{du_steady_2024}%
  \BibitemOpen
  \bibfield  {author} {\bibinfo {author} {\bibfnamefont {H.}~\bibnamefont {Du}}, \bibinfo {author} {\bibfnamefont {D.}~\bibnamefont {Song}}, \bibinfo {author} {\bibfnamefont {W.}~\bibnamefont {Wang}}, \bibinfo {author} {\bibfnamefont {S.}~\bibnamefont {Zhang}}, \bibinfo {author} {\bibfnamefont {N.}~\bibnamefont {Wang}}, \bibinfo {author} {\bibfnamefont {F.}~\bibnamefont {Zheng}}, \bibinfo {author} {\bibfnamefont {M.}~\bibnamefont {Tian}}, \bibinfo {author} {\bibfnamefont {R.}~\bibnamefont {Dunin-Borkowski}}, \ and\ \bibinfo {author} {\bibfnamefont {J.}~\bibnamefont {Zang}},\ }\bibfield  {title} {\enquote {\bibinfo {title} {Steady motion of 80-nm-size skyrmions in a 100-nm-wide track},}\ }\href {\doibase 10.21203/rs.3.rs-4053583/v1} {\  (\bibinfo {year} {2024}),\ 10.21203/rs.3.rs-4053583/v1}\BibitemShut {NoStop}%
\bibitem [{\citenamefont {Liu}\ \emph {et~al.}(2024)\citenamefont {Liu}, \citenamefont {Wang}, \citenamefont {He}, \citenamefont {Zhang}, \citenamefont {Zhang}, \citenamefont {Yuan}, \citenamefont {Hou}, \citenamefont {Qin}, \citenamefont {Xu}, \citenamefont {Gao}, \citenamefont {Peng}, \citenamefont {Liu}, \citenamefont {Qiu}, \citenamefont {Liu},\ and\ \citenamefont {Zhang}}]{liu_strain-induced_2024}%
  \BibitemOpen
  \bibfield  {author} {\bibinfo {author} {\bibfnamefont {C.}~\bibnamefont {Liu}}, \bibinfo {author} {\bibfnamefont {J.}~\bibnamefont {Wang}}, \bibinfo {author} {\bibfnamefont {W.}~\bibnamefont {He}}, \bibinfo {author} {\bibfnamefont {C.}~\bibnamefont {Zhang}}, \bibinfo {author} {\bibfnamefont {S.}~\bibnamefont {Zhang}}, \bibinfo {author} {\bibfnamefont {S.}~\bibnamefont {Yuan}}, \bibinfo {author} {\bibfnamefont {Z.}~\bibnamefont {Hou}}, \bibinfo {author} {\bibfnamefont {M.}~\bibnamefont {Qin}}, \bibinfo {author} {\bibfnamefont {Y.}~\bibnamefont {Xu}}, \bibinfo {author} {\bibfnamefont {X.}~\bibnamefont {Gao}}, \bibinfo {author} {\bibfnamefont {Y.}~\bibnamefont {Peng}}, \bibinfo {author} {\bibfnamefont {K.}~\bibnamefont {Liu}}, \bibinfo {author} {\bibfnamefont {Z.~Q.}\ \bibnamefont {Qiu}}, \bibinfo {author} {\bibfnamefont {J.-M.}\ \bibnamefont {Liu}}, \ and\ \bibinfo {author} {\bibfnamefont {X.}~\bibnamefont {Zhang}},\ }\bibfield  {title} {\enquote {\bibinfo {title} {Strain-induced reversible motion of
  skyrmions at room temperature},}\ }\href {\doibase 10.1021/acsnano.3c09090} {\bibfield  {journal} {\bibinfo  {journal} {ACS Nano}\ }\textbf {\bibinfo {volume} {18}},\ \bibinfo {pages} {761--769} (\bibinfo {year} {2024})}\BibitemShut {NoStop}%
\bibitem [{\citenamefont {Xing}\ and\ \citenamefont {Zhou}(2022)}]{xing_skyrmion_2022}%
  \BibitemOpen
  \bibfield  {author} {\bibinfo {author} {\bibfnamefont {X.}~\bibnamefont {Xing}}\ and\ \bibinfo {author} {\bibfnamefont {Y.}~\bibnamefont {Zhou}},\ }\bibfield  {title} {\enquote {\bibinfo {title} {Skyrmion motion and partitioning of domain wall velocity driven by repulsive interactions},}\ }\href {\doibase 10.1038/s42005-022-01020-z} {\bibfield  {journal} {\bibinfo  {journal} {Commun. Phys.}\ }\textbf {\bibinfo {volume} {5}},\ \bibinfo {pages} {1--11} (\bibinfo {year} {2022})}\BibitemShut {NoStop}%
\bibitem [{\citenamefont {Reichhardt}\ and\ \citenamefont {Reichhardt}(2015)}]{reichhardt_shapiro_2015}%
  \BibitemOpen
  \bibfield  {author} {\bibinfo {author} {\bibfnamefont {C.}~\bibnamefont {Reichhardt}}\ and\ \bibinfo {author} {\bibfnamefont {C.~J.~Olson}\ \bibnamefont {Reichhardt}},\ }\bibfield  {title} {\enquote {\bibinfo {title} {Shapiro steps for skyrmion motion on a washboard potential with longitudinal and transverse ac drives},}\ }\href {\doibase 10.1103/PhysRevB.92.224432} {\bibfield  {journal} {\bibinfo  {journal} {Phys. Rev. B}\ }\textbf {\bibinfo {volume} {92}},\ \bibinfo {pages} {224432} (\bibinfo {year} {2015})}\BibitemShut {NoStop}%
\bibitem [{\citenamefont {Reichhardt}\ and\ \citenamefont {Reichhardt}(2017{\natexlab{b}})}]{reichhardt_shapiro_2017}%
  \BibitemOpen
  \bibfield  {author} {\bibinfo {author} {\bibfnamefont {C.}~\bibnamefont {Reichhardt}}\ and\ \bibinfo {author} {\bibfnamefont {C.~J.~Olson}\ \bibnamefont {Reichhardt}},\ }\bibfield  {title} {\enquote {\bibinfo {title} {Shapiro spikes and negative mobility for skyrmion motion on quasi-one-dimensional periodic substrates},}\ }\href {\doibase 10.1103/PhysRevB.95.014412} {\bibfield  {journal} {\bibinfo  {journal} {Phys. Rev. B}\ }\textbf {\bibinfo {volume} {95}},\ \bibinfo {pages} {014412} (\bibinfo {year} {2017}{\natexlab{b}})}\BibitemShut {NoStop}%
\bibitem [{\citenamefont {Vizarim}\ \emph {et~al.}(2020{\natexlab{b}})\citenamefont {Vizarim}, \citenamefont {Reichhardt}, \citenamefont {Venegas},\ and\ \citenamefont {Reichhardt}}]{vizarim_shapiro_2020}%
  \BibitemOpen
  \bibfield  {author} {\bibinfo {author} {\bibfnamefont {N.~P.}\ \bibnamefont {Vizarim}}, \bibinfo {author} {\bibfnamefont {C.}~\bibnamefont {Reichhardt}}, \bibinfo {author} {\bibfnamefont {P.~A.}\ \bibnamefont {Venegas}}, \ and\ \bibinfo {author} {\bibfnamefont {C.~J.~O.}\ \bibnamefont {Reichhardt}},\ }\bibfield  {title} {\enquote {\bibinfo {title} {Shapiro steps and nonlinear skyrmion {Hall} angles for dc and ac driven skyrmions on a two-dimensional periodic substrate},}\ }\href {\doibase 10.1103/PhysRevB.102.104413} {\bibfield  {journal} {\bibinfo  {journal} {Phys. Rev. B}\ }\textbf {\bibinfo {volume} {102}},\ \bibinfo {pages} {104413} (\bibinfo {year} {2020}{\natexlab{b}})}\BibitemShut {NoStop}%
\bibitem [{\citenamefont {Evans}(2018)}]{evans_atomistic_2018}%
  \BibitemOpen
  \bibfield  {author} {\bibinfo {author} {\bibfnamefont {R.~F.~L.}\ \bibnamefont {Evans}},\ }\bibfield  {title} {\enquote {\bibinfo {title} {Atomistic {Spin} {Dynamics}},}\ }in\ \href {\doibase 10.1007/978-3-319-50257-1_147-1} {\emph {\bibinfo {booktitle} {Handbook of {Materials} {Modeling}: {Applications}: {Current} and {Emerging} {Materials}}}},\ \bibinfo {editor} {edited by\ \bibinfo {editor} {\bibfnamefont {W.}~\bibnamefont {Andreoni}}\ and\ \bibinfo {editor} {\bibfnamefont {S.}~\bibnamefont {Yip}}}\ (\bibinfo  {publisher} {Springer International Publishing},\ \bibinfo {year} {2018})\ pp.\ \bibinfo {pages} {1--23}\BibitemShut {NoStop}%
\bibitem [{\citenamefont {Iwasaki}\ \emph {et~al.}(2013{\natexlab{b}})\citenamefont {Iwasaki}, \citenamefont {Mochizuki},\ and\ \citenamefont {Nagaosa}}]{iwasaki_current-induced_2013}%
  \BibitemOpen
  \bibfield  {author} {\bibinfo {author} {\bibfnamefont {J.}~\bibnamefont {Iwasaki}}, \bibinfo {author} {\bibfnamefont {M.}~\bibnamefont {Mochizuki}}, \ and\ \bibinfo {author} {\bibfnamefont {N.}~\bibnamefont {Nagaosa}},\ }\bibfield  {title} {\enquote {\bibinfo {title} {Current-induced skyrmion dynamics in constricted geometries},}\ }\href {\doibase 10.1038/nnano.2013.176} {\bibfield  {journal} {\bibinfo  {journal} {Nature Nanotechnol.}\ }\textbf {\bibinfo {volume} {8}},\ \bibinfo {pages} {742--747} (\bibinfo {year} {2013}{\natexlab{b}})}\BibitemShut {NoStop}%
\bibitem [{\citenamefont {Paul}\ \emph {et~al.}(2020)\citenamefont {Paul}, \citenamefont {Haldar}, \citenamefont {von Malottki},\ and\ \citenamefont {Heinze}}]{paul_role_2020}%
  \BibitemOpen
  \bibfield  {author} {\bibinfo {author} {\bibfnamefont {S.}~\bibnamefont {Paul}}, \bibinfo {author} {\bibfnamefont {S.}~\bibnamefont {Haldar}}, \bibinfo {author} {\bibfnamefont {S.}~\bibnamefont {von Malottki}}, \ and\ \bibinfo {author} {\bibfnamefont {S.}~\bibnamefont {Heinze}},\ }\bibfield  {title} {\enquote {\bibinfo {title} {Role of higher-order exchange interactions for skyrmion stability},}\ }\href {\doibase 10.1038/s41467-020-18473-x} {\bibfield  {journal} {\bibinfo  {journal} {Nature Commun.}\ }\textbf {\bibinfo {volume} {11}},\ \bibinfo {pages} {4756} (\bibinfo {year} {2020})}\BibitemShut {NoStop}%
\bibitem [{\citenamefont {Seki}\ and\ \citenamefont {Mochizuki}(2016)}]{seki_skyrmions_2016}%
  \BibitemOpen
  \bibfield  {author} {\bibinfo {author} {\bibfnamefont {S.}~\bibnamefont {Seki}}\ and\ \bibinfo {author} {\bibfnamefont {M.}~\bibnamefont {Mochizuki}},\ }\href {\doibase 10.1007/978-3-319-24651-2} {\emph {\bibinfo {title} {Skyrmions in {Magnetic} {Materials}}}}\ (\bibinfo  {publisher} {Springer International Publishing},\ \bibinfo {year} {2016})\BibitemShut {NoStop}%
\bibitem [{\citenamefont {Slonczewski}(1972)}]{slonczewski_dynamics_1972}%
  \BibitemOpen
  \bibfield  {author} {\bibinfo {author} {\bibfnamefont {J.~C.}\ \bibnamefont {Slonczewski}},\ }\bibfield  {title} {\enquote {\bibinfo {title} {Dynamics of magnetic domain walls},}\ }\href {\doibase 10.1063/1.3699416} {\bibfield  {journal} {\bibinfo  {journal} {AIP Conf. Proc.}\ }\textbf {\bibinfo {volume} {5}},\ \bibinfo {pages} {170--174} (\bibinfo {year} {1972})}\BibitemShut {NoStop}%
\bibitem [{\citenamefont {Gilbert}(2004)}]{gilbert_phenomenological_2004}%
  \BibitemOpen
  \bibfield  {author} {\bibinfo {author} {\bibfnamefont {T.~L.}\ \bibnamefont {Gilbert}},\ }\bibfield  {title} {\enquote {\bibinfo {title} {A phenomenological theory of damping in ferromagnetic materials},}\ }\href {\doibase 10.1109/TMAG.2004.836740} {\bibfield  {journal} {\bibinfo  {journal} {IEEE Trans. Mag.}\ }\textbf {\bibinfo {volume} {40}},\ \bibinfo {pages} {3443--3449} (\bibinfo {year} {2004})}\BibitemShut {NoStop}%
\bibitem [{\citenamefont {Zang}\ \emph {et~al.}(2011)\citenamefont {Zang}, \citenamefont {Mostovoy}, \citenamefont {Han},\ and\ \citenamefont {Nagaosa}}]{zang_dynamics_2011}%
  \BibitemOpen
  \bibfield  {author} {\bibinfo {author} {\bibfnamefont {J.}~\bibnamefont {Zang}}, \bibinfo {author} {\bibfnamefont {M.}~\bibnamefont {Mostovoy}}, \bibinfo {author} {\bibfnamefont {J.~H.}\ \bibnamefont {Han}}, \ and\ \bibinfo {author} {\bibfnamefont {N.}~\bibnamefont {Nagaosa}},\ }\bibfield  {title} {\enquote {\bibinfo {title} {Dynamics of skyrmion crystals in metallic thin films},}\ }\href {\doibase 10.1103/PhysRevLett.107.136804} {\bibfield  {journal} {\bibinfo  {journal} {Phys. Rev. Lett.}\ }\textbf {\bibinfo {volume} {107}},\ \bibinfo {pages} {136804} (\bibinfo {year} {2011})}\BibitemShut {NoStop}%
\bibitem [{\citenamefont {Schulz}\ \emph {et~al.}(2012)\citenamefont {Schulz}, \citenamefont {Ritz}, \citenamefont {Bauer}, \citenamefont {Halder}, \citenamefont {Wagner}, \citenamefont {Franz}, \citenamefont {Pfleiderer}, \citenamefont {Everschor}, \citenamefont {Garst},\ and\ \citenamefont {Rosch}}]{schulz_emergent_2012}%
  \BibitemOpen
  \bibfield  {author} {\bibinfo {author} {\bibfnamefont {T.}~\bibnamefont {Schulz}}, \bibinfo {author} {\bibfnamefont {R.}~\bibnamefont {Ritz}}, \bibinfo {author} {\bibfnamefont {A.}~\bibnamefont {Bauer}}, \bibinfo {author} {\bibfnamefont {M.}~\bibnamefont {Halder}}, \bibinfo {author} {\bibfnamefont {M.}~\bibnamefont {Wagner}}, \bibinfo {author} {\bibfnamefont {C.}~\bibnamefont {Franz}}, \bibinfo {author} {\bibfnamefont {C.}~\bibnamefont {Pfleiderer}}, \bibinfo {author} {\bibfnamefont {K.}~\bibnamefont {Everschor}}, \bibinfo {author} {\bibfnamefont {M.}~\bibnamefont {Garst}}, \ and\ \bibinfo {author} {\bibfnamefont {A.}~\bibnamefont {Rosch}},\ }\bibfield  {title} {\enquote {\bibinfo {title} {Emergent electrodynamics of skyrmions in a chiral magnet},}\ }\href {\doibase 10.1038/NPHYS2231} {\bibfield  {journal} {\bibinfo  {journal} {Nature Phys.}\ }\textbf {\bibinfo {volume} {8}},\ \bibinfo {pages} {301--304} (\bibinfo {year} {2012})}\BibitemShut {NoStop}%
\bibitem [{\citenamefont {Boulle}\ \emph {et~al.}(2016)\citenamefont {Boulle}, \citenamefont {Vogel}, \citenamefont {Yang}, \citenamefont {Pizzini}, \citenamefont {Chaves}, \citenamefont {Locatelli}, \citenamefont {Mente{\c s}}, \citenamefont {Sala}, \citenamefont {Buda-Prejbeanu}, \citenamefont {Klein}, \citenamefont {Belmeguenai}, \citenamefont {Roussign{\' e}}, \citenamefont {Stashkevich}, \citenamefont {Ch{\' e}rif}, \citenamefont {Aballe}, \citenamefont {Foerster}, \citenamefont {Chshiev}, \citenamefont {Auffret}, \citenamefont {Miron},\ and\ \citenamefont {Gaudin}}]{boulle_room-temperature_2016}%
  \BibitemOpen
  \bibfield  {author} {\bibinfo {author} {\bibfnamefont {O.}~\bibnamefont {Boulle}}, \bibinfo {author} {\bibfnamefont {J.}~\bibnamefont {Vogel}}, \bibinfo {author} {\bibfnamefont {H.}~\bibnamefont {Yang}}, \bibinfo {author} {\bibfnamefont {S.}~\bibnamefont {Pizzini}}, \bibinfo {author} {\bibfnamefont {D.~de~Souza}\ \bibnamefont {Chaves}}, \bibinfo {author} {\bibfnamefont {A.}~\bibnamefont {Locatelli}}, \bibinfo {author} {\bibfnamefont {T.~O.}\ \bibnamefont {Mente{\c s}}}, \bibinfo {author} {\bibfnamefont {A.}~\bibnamefont {Sala}}, \bibinfo {author} {\bibfnamefont {L.~D.}\ \bibnamefont {Buda-Prejbeanu}}, \bibinfo {author} {\bibfnamefont {O.}~\bibnamefont {Klein}}, \bibinfo {author} {\bibfnamefont {M.}~\bibnamefont {Belmeguenai}}, \bibinfo {author} {\bibfnamefont {Y.}~\bibnamefont {Roussign{\' e}}}, \bibinfo {author} {\bibfnamefont {A.}~\bibnamefont {Stashkevich}}, \bibinfo {author} {\bibfnamefont {S.~M.}\ \bibnamefont {Ch{\' e}rif}}, \bibinfo {author} {\bibfnamefont {L.}~\bibnamefont {Aballe}}, \bibinfo
  {author} {\bibfnamefont {M.}~\bibnamefont {Foerster}}, \bibinfo {author} {\bibfnamefont {M.}~\bibnamefont {Chshiev}}, \bibinfo {author} {\bibfnamefont {S.}~\bibnamefont {Auffret}}, \bibinfo {author} {\bibfnamefont {I.~M.}\ \bibnamefont {Miron}}, \ and\ \bibinfo {author} {\bibfnamefont {G.}~\bibnamefont {Gaudin}},\ }\bibfield  {title} {\enquote {\bibinfo {title} {Room-temperature chiral magnetic skyrmions in ultrathin magnetic nanostructures},}\ }\href {\doibase 10.1038/nnano.2015.315} {\bibfield  {journal} {\bibinfo  {journal} {Nature Nanotechnol.}\ }\textbf {\bibinfo {volume} {11}},\ \bibinfo {pages} {449--454} (\bibinfo {year} {2016})}\BibitemShut {NoStop}%
\bibitem [{\citenamefont {Juge}\ \emph {et~al.}(2021)\citenamefont {Juge}, \citenamefont {Bairagi}, \citenamefont {Rana}, \citenamefont {Vogel}, \citenamefont {Sall}, \citenamefont {Mailly}, \citenamefont {Pham}, \citenamefont {Zhang}, \citenamefont {Sisodia}, \citenamefont {Foerster}, \citenamefont {Aballe}, \citenamefont {Belmeguenai}, \citenamefont {Roussign{\' e}}, \citenamefont {Auffret}, \citenamefont {Buda-Prejbeanu}, \citenamefont {Gaudin}, \citenamefont {Ravelosona},\ and\ \citenamefont {Boulle}}]{juge_helium_2021}%
  \BibitemOpen
  \bibfield  {author} {\bibinfo {author} {\bibfnamefont {R.}~\bibnamefont {Juge}}, \bibinfo {author} {\bibfnamefont {K.}~\bibnamefont {Bairagi}}, \bibinfo {author} {\bibfnamefont {K.~G.}\ \bibnamefont {Rana}}, \bibinfo {author} {\bibfnamefont {J.}~\bibnamefont {Vogel}}, \bibinfo {author} {\bibfnamefont {M.}~\bibnamefont {Sall}}, \bibinfo {author} {\bibfnamefont {D.}~\bibnamefont {Mailly}}, \bibinfo {author} {\bibfnamefont {V.~T.}\ \bibnamefont {Pham}}, \bibinfo {author} {\bibfnamefont {Q.}~\bibnamefont {Zhang}}, \bibinfo {author} {\bibfnamefont {N.}~\bibnamefont {Sisodia}}, \bibinfo {author} {\bibfnamefont {M.}~\bibnamefont {Foerster}}, \bibinfo {author} {\bibfnamefont {L.}~\bibnamefont {Aballe}}, \bibinfo {author} {\bibfnamefont {M.}~\bibnamefont {Belmeguenai}}, \bibinfo {author} {\bibfnamefont {Y.}~\bibnamefont {Roussign{\' e}}}, \bibinfo {author} {\bibfnamefont {S.}~\bibnamefont {Auffret}}, \bibinfo {author} {\bibfnamefont {L.~D.}\ \bibnamefont {Buda-Prejbeanu}}, \bibinfo {author} {\bibfnamefont
  {G.}~\bibnamefont {Gaudin}}, \bibinfo {author} {\bibfnamefont {D.}~\bibnamefont {Ravelosona}}, \ and\ \bibinfo {author} {\bibfnamefont {O.}~\bibnamefont {Boulle}},\ }\bibfield  {title} {\enquote {\bibinfo {title} {Helium ions put magnetic skyrmions on the track},}\ }\href {\doibase 10.1021/acs.nanolett.1c00136} {\bibfield  {journal} {\bibinfo  {journal} {Nano Lett.}\ }\textbf {\bibinfo {volume} {21}},\ \bibinfo {pages} {2989--2996} (\bibinfo {year} {2021})}\BibitemShut {NoStop}%
\bibitem [{\citenamefont {Souza}\ \emph {et~al.}(2022)\citenamefont {Souza}, \citenamefont {Vizarim}, \citenamefont {Reichhardt}, \citenamefont {Reichhardt},\ and\ \citenamefont {Venegas}}]{souza_clogging_2022}%
  \BibitemOpen
  \bibfield  {author} {\bibinfo {author} {\bibfnamefont {J.~C.~Bellizotti}\ \bibnamefont {Souza}}, \bibinfo {author} {\bibfnamefont {N.~P.}\ \bibnamefont {Vizarim}}, \bibinfo {author} {\bibfnamefont {C.~J.~O.}\ \bibnamefont {Reichhardt}}, \bibinfo {author} {\bibfnamefont {C.}~\bibnamefont {Reichhardt}}, \ and\ \bibinfo {author} {\bibfnamefont {P.~A.}\ \bibnamefont {Venegas}},\ }\bibfield  {title} {\enquote {\bibinfo {title} {Clogging, diode and collective effects of skyrmions in funnel geometries},}\ }\href@noop {} {\bibfield  {journal} {\bibinfo  {journal} {New J. Phys.}\ }\textbf {\bibinfo {volume} {24}},\ \bibinfo {pages} {103030} (\bibinfo {year} {2022})}\BibitemShut {NoStop}%
\bibitem [{\citenamefont {Landau}\ and\ \citenamefont {Lifshitz}(1976)}]{Landau76}%
  \BibitemOpen
  \bibfield  {author} {\bibinfo {author} {\bibfnamefont {L.~D.}\ \bibnamefont {Landau}}\ and\ \bibinfo {author} {\bibfnamefont {E.~M.}\ \bibnamefont {Lifshitz}},\ }\href@noop {} {\emph {\bibinfo {title} {Mechanics}}},\ \bibinfo {edition} {3rd}\ ed.\ (\bibinfo  {publisher} {Pergamon},\ \bibinfo {year} {1976})\BibitemShut {NoStop}%
\bibitem [{\citenamefont {Reichhardt}\ and\ \citenamefont {Olson}(2002)}]{Reichhardt02a}%
  \BibitemOpen
  \bibfield  {author} {\bibinfo {author} {\bibfnamefont {C.}~\bibnamefont {Reichhardt}}\ and\ \bibinfo {author} {\bibfnamefont {C.~J.}\ \bibnamefont {Olson}},\ }\bibfield  {title} {\enquote {\bibinfo {title} {Transverse phase locking for vortex motion in square and triangular pinning arrays},}\ }\href {\doibase 10.1103/PhysRevB.65.174523} {\bibfield  {journal} {\bibinfo  {journal} {Phys. Rev. B}\ }\textbf {\bibinfo {volume} {65}},\ \bibinfo {pages} {174523} (\bibinfo {year} {2002})}\BibitemShut {NoStop}%
\end{thebibliography}%

\end{document}